\iftrue
\def\targetjournal{aa}
\documentclass[structabstract,longauth]{aa}   
\else
\def\targetjournal{apj}
\documentclass[preprint]{aastex}     
\fi

\usepackage{natbib}
\usepackage{color}
\usepackage{graphicx}
\usepackage{latexsym}
\usepackage{float}
\usepackage{amsmath}
\usepackage{amssymb}
\usepackage{multirow}
\usepackage{subfigure}
\usepackage{txfonts}   
\usepackage{ifthen}
\usepackage{setspace}
\usepackage{supertabular}
\usepackage{lscape}

\bibpunct{(}{)}{;}{a}{}{,} 
\renewcommand{\today}{\number\day\space\ifcase\month\or
  January\or February\or March\or April\or May\or June\or
  July\or August\or September\or October\or November\or December\fi
  \space\number\year}

\newcommand{\ligodoc}{{LIGO-P10}{00061}-{v19}} 

\begin{document}

\title{Implementation and testing of the first prompt search for gravitational wave transients with electromagnetic counterparts \vspace*{-0.1in} }
\ifthenelse{\equal{\targetjournal}{aa}}{}{
\slugcomment{ \textcolor{red}{{\large \ligodoc}}  Dated: \today }
}

\iffalse
  
\author{J.~Abadie\altaffilmark{29},
B.~P.~Abbott\altaffilmark{29},
R.~Abbott\altaffilmark{29},
F.~Acernese\altaffilmark{19ac},
R.~Adhikari\altaffilmark{29},
P.~Ajith\altaffilmark{2},
B.~Allen\altaffilmark{2,78},
G.~Allen\altaffilmark{52},
R.~S.~Amin\altaffilmark{34},
S.~B.~Anderson\altaffilmark{29}
% ...
}
% ...
\altaffiltext{2}{Albert-Einstein-Institut, Max-Planck-Institut f\"ur Gravitationsphysik, D-30167 Hannover, Germany}
% ...
\altaffiltext{19}{INFN, Sezione di Napoli $^a$; Universit\`a di Napoli 'Federico II'$^b$ Complesso Universitario di Monte S.Angelo, I-80126 Napoli; Universit\`a di Salerno, Fisciano, I-84084 Salerno$^c$, Italy}
% ...
\altaffiltext{29}{LIGO - California Institute of Technology, Pasadena, CA  91125, USA }
% ...
\altaffiltext{34}{Louisiana State University, Baton Rouge, LA  70803, USA }
% ...
\altaffiltext{52}{Stanford University, Stanford, CA  94305, USA }
% ...
\altaffiltext{78}{University of Wisconsin--Milwaukee, Milwaukee, WI  53201, USA }
% ...

\else
  
\author{
\scriptsize{
\begin{spacing}{0.3}
The LIGO Scientific Collaboration and Virgo Collaboration: J.~Abadie\inst{\ref{inst1}} \and 
B.~P.~Abbott\inst{\ref{inst1}} \and 
R.~Abbott\inst{\ref{inst1}} \and 
T.~D.~Abbott\inst{\ref{inst2}} \and 
M.~Abernathy\inst{\ref{inst3}} \and 
T.~Accadia\inst{\ref{inst4}} \and 
F.~Acernese\inst{\ref{inst5}ac} \and 
C.~Adams\inst{\ref{inst6}} \and 
R.~Adhikari\inst{\ref{inst1}} \and 
C.~Affeldt\inst{\ref{inst7},\ref{inst8}} \and 
P.~Ajith\inst{\ref{inst1}} \and 
B.~Allen\inst{\ref{inst7},\ref{inst9},\ref{inst8}} \and 
G.~S.~Allen\inst{\ref{inst10}} \and 
E.~Amador~Ceron\inst{\ref{inst9}} \and 
D.~Amariutei\inst{\ref{inst11}} \and 
R.~S.~Amin\inst{\ref{inst12}} \and 
S.~B.~Anderson\inst{\ref{inst1}} \and 
W.~G.~Anderson\inst{\ref{inst9}} \and 
K.~Arai\inst{\ref{inst1}} \and 
M.~A.~Arain\inst{\ref{inst11}} \and 
M.~C.~Araya\inst{\ref{inst1}} \and 
S.~M.~Aston\inst{\ref{inst13}} \and 
P.~Astone\inst{\ref{inst14}a} \and 
D.~Atkinson\inst{\ref{inst15}} \and 
P.~Aufmuth\inst{\ref{inst8},\ref{inst7}} \and 
C.~Aulbert\inst{\ref{inst7},\ref{inst8}} \and 
B.~E.~Aylott\inst{\ref{inst13}} \and 
S.~Babak\inst{\ref{inst16}} \and 
P.~Baker\inst{\ref{inst17}} \and 
G.~Ballardin\inst{\ref{inst18}} \and 
S.~Ballmer\inst{\ref{inst19}} \and 
D.~Barker\inst{\ref{inst15}} \and 
F.~Barone\inst{\ref{inst5}ac} \and 
B.~Barr\inst{\ref{inst3}} \and 
P.~Barriga\inst{\ref{inst20}} \and 
L.~Barsotti\inst{\ref{inst21}} \and 
M.~Barsuglia\inst{\ref{inst22}} \and 
M.~A.~Barton\inst{\ref{inst15}} \and 
I.~Bartos\inst{\ref{inst23}} \and 
R.~Bassiri\inst{\ref{inst3}} \and 
M.~Bastarrika\inst{\ref{inst3}} \and 
A.~Basti\inst{\ref{inst24}ab} \and 
J.~Batch\inst{\ref{inst15}} \and 
J.~Bauchrowitz\inst{\ref{inst7},\ref{inst8}} \and 
Th.~S.~Bauer\inst{\ref{inst25}a} \and 
M.~Bebronne\inst{\ref{inst4}} \and 
B.~Behnke\inst{\ref{inst16}} \and 
M.G.~Beker\inst{\ref{inst25}a} \and 
A.~S.~Bell\inst{\ref{inst3}} \and 
A.~Belletoile\inst{\ref{inst4}} \and 
I.~Belopolski\inst{\ref{inst23}} \and 
M.~Benacquista\inst{\ref{inst26}} \and 
J.~M.~Berliner\inst{\ref{inst15}} \and 
A.~Bertolini\inst{\ref{inst7},\ref{inst8}} \and 
J.~Betzwieser\inst{\ref{inst1}} \and 
N.~Beveridge\inst{\ref{inst3}} \and 
P.~T.~Beyersdorf\inst{\ref{inst27}} \and 
I.~A.~Bilenko\inst{\ref{inst28}} \and 
G.~Billingsley\inst{\ref{inst1}} \and 
J.~Birch\inst{\ref{inst6}} \and 
R.~Biswas\inst{\ref{inst26}} \and 
M.~Bitossi\inst{\ref{inst24}a} \and 
M.~A.~Bizouard\inst{\ref{inst29}a} \and 
E.~Black\inst{\ref{inst1}} \and 
J.~K.~Blackburn\inst{\ref{inst1}} \and 
L.~Blackburn\inst{\ref{inst30}} \and 
D.~Blair\inst{\ref{inst20}} \and 
B.~Bland\inst{\ref{inst15}} \and 
M.~Blom\inst{\ref{inst25}a} \and 
O.~Bock\inst{\ref{inst7},\ref{inst8}} \and 
T.~P.~Bodiya\inst{\ref{inst21}} \and 
C.~Bogan\inst{\ref{inst7},\ref{inst8}} \and 
R.~Bondarescu\inst{\ref{inst31}} \and 
F.~Bondu\inst{\ref{inst32}b} \and 
L.~Bonelli\inst{\ref{inst24}ab} \and 
R.~Bonnand\inst{\ref{inst33}} \and 
R.~Bork\inst{\ref{inst1}} \and 
M.~Born\inst{\ref{inst7},\ref{inst8}} \and 
V.~Boschi\inst{\ref{inst24}a} \and 
S.~Bose\inst{\ref{inst34}} \and 
L.~Bosi\inst{\ref{inst35}a} \and 
B. ~Bouhou\inst{\ref{inst22}} \and 
S.~Braccini\inst{\ref{inst24}a} \and 
C.~Bradaschia\inst{\ref{inst24}a} \and 
P.~R.~Brady\inst{\ref{inst9}} \and 
V.~B.~Braginsky\inst{\ref{inst28}} \and 
M.~Branchesi\inst{\ref{inst36}ab} \and 
J.~E.~Brau\inst{\ref{inst37}} \and 
J.~Breyer\inst{\ref{inst7},\ref{inst8}} \and 
T.~Briant\inst{\ref{inst38}} \and 
D.~O.~Bridges\inst{\ref{inst6}} \and 
A.~Brillet\inst{\ref{inst32}a} \and 
M.~Brinkmann\inst{\ref{inst7},\ref{inst8}} \and 
V.~Brisson\inst{\ref{inst29}a} \and 
M.~Britzger\inst{\ref{inst7},\ref{inst8}} \and 
A.~F.~Brooks\inst{\ref{inst1}} \and 
D.~A.~Brown\inst{\ref{inst19}} \and 
A.~Brummit\inst{\ref{inst39}} \and 
T.~Bulik\inst{\ref{inst40}bc} \and 
H.~J.~Bulten\inst{\ref{inst25}ab} \and 
A.~Buonanno\inst{\ref{inst41}} \and 
J.~Burguet--Castell\inst{\ref{inst9}} \and 
O.~Burmeister\inst{\ref{inst7},\ref{inst8}} \and 
D.~Buskulic\inst{\ref{inst4}} \and 
C.~Buy\inst{\ref{inst22}} \and 
R.~L.~Byer\inst{\ref{inst10}} \and 
L.~Cadonati\inst{\ref{inst42}} \and 
G.~Cagnoli\inst{\ref{inst36}a} \and 
E.~Calloni\inst{\ref{inst5}ab} \and 
J.~B.~Camp\inst{\ref{inst30}} \and 
P.~Campsie\inst{\ref{inst3}} \and 
J.~Cannizzo\inst{\ref{inst30}} \and 
K.~Cannon\inst{\ref{inst44}} \and 
B.~Canuel\inst{\ref{inst18}} \and 
J.~Cao\inst{\ref{inst45}} \and 
C.~D.~Capano\inst{\ref{inst19}} \and 
F.~Carbognani\inst{\ref{inst18}} \and 
S.~Caride\inst{\ref{inst46}} \and 
S.~Caudill\inst{\ref{inst12}} \and 
M.~Cavagli\`a\inst{\ref{inst43}} \and 
F.~Cavalier\inst{\ref{inst29}a} \and 
R.~Cavalieri\inst{\ref{inst18}} \and 
G.~Cella\inst{\ref{inst24}a} \and 
C.~Cepeda\inst{\ref{inst1}} \and 
E.~Cesarini\inst{\ref{inst36}b} \and 
O.~Chaibi\inst{\ref{inst32}a} \and 
T.~Chalermsongsak\inst{\ref{inst1}} \and 
E.~Chalkley\inst{\ref{inst13}} \and 
P.~Charlton\inst{\ref{inst47}} \and 
E.~Chassande-Mottin\inst{\ref{inst22}} \and 
S.~Chelkowski\inst{\ref{inst13}} \and 
Y.~Chen\inst{\ref{inst48}} \and 
A.~Chincarini\inst{\ref{inst49}} \and 
A.~Chiummo\inst{\ref{inst18}} \and 
H.~Cho\inst{\ref{inst50}} \and 
N.~Christensen\inst{\ref{inst51}} \and 
S.~S.~Y.~Chua\inst{\ref{inst52}} \and 
C.~T.~Y.~Chung\inst{\ref{inst53}} \and 
S.~Chung\inst{\ref{inst20}} \and 
G.~Ciani\inst{\ref{inst11}} \and 
F.~Clara\inst{\ref{inst15}} \and 
D.~E.~Clark\inst{\ref{inst10}} \and 
J.~Clark\inst{\ref{inst54}} \and 
J.~H.~Clayton\inst{\ref{inst9}} \and 
F.~Cleva\inst{\ref{inst32}a} \and 
E.~Coccia\inst{\ref{inst55}ab} \and 
P.-F.~Cohadon\inst{\ref{inst38}} \and 
C.~N.~Colacino\inst{\ref{inst24}ab} \and 
J.~Colas\inst{\ref{inst18}} \and 
A.~Colla\inst{\ref{inst14}ab} \and 
M.~Colombini\inst{\ref{inst14}b} \and 
A.~Conte\inst{\ref{inst14}ab} \and 
R.~Conte\inst{\ref{inst56}} \and 
D.~Cook\inst{\ref{inst15}} \and 
T.~R.~Corbitt\inst{\ref{inst21}} \and 
M.~Cordier\inst{\ref{inst27}} \and 
N.~Cornish\inst{\ref{inst17}} \and 
A.~Corsi\inst{\ref{inst1}} \and 
C.~A.~Costa\inst{\ref{inst12}} \and 
M.~Coughlin\inst{\ref{inst51}} \and 
J.-P.~Coulon\inst{\ref{inst32}a} \and 
P.~Couvares\inst{\ref{inst19}} \and 
D.~M.~Coward\inst{\ref{inst20}} \and 
D.~C.~Coyne\inst{\ref{inst1}} \and 
J.~D.~E.~Creighton\inst{\ref{inst9}} \and 
T.~D.~Creighton\inst{\ref{inst26}} \and 
A.~M.~Cruise\inst{\ref{inst13}} \and 
A.~Cumming\inst{\ref{inst3}} \and 
L.~Cunningham\inst{\ref{inst3}} \and 
E.~Cuoco\inst{\ref{inst18}} \and 
R.~M.~Cutler\inst{\ref{inst13}} \and 
K.~Dahl\inst{\ref{inst7},\ref{inst8}} \and 
S.~L.~Danilishin\inst{\ref{inst28}} \and 
R.~Dannenberg\inst{\ref{inst1}} \and 
S.~D'Antonio\inst{\ref{inst55}a} \and 
K.~Danzmann\inst{\ref{inst7},\ref{inst8}} \and 
V.~Dattilo\inst{\ref{inst18}} \and 
B.~Daudert\inst{\ref{inst1}} \and 
H.~Daveloza\inst{\ref{inst26}} \and 
M.~Davier\inst{\ref{inst29}a} \and 
G.~Davies\inst{\ref{inst54}} \and 
E.~J.~Daw\inst{\ref{inst57}} \and 
R.~Day\inst{\ref{inst18}} \and 
T.~Dayanga\inst{\ref{inst34}} \and 
R.~De~Rosa\inst{\ref{inst5}ab} \and 
D.~DeBra\inst{\ref{inst10}} \and 
G.~Debreczeni\inst{\ref{inst58}} \and 
J.~Degallaix\inst{\ref{inst7},\ref{inst8}} \and 
W.~Del~Pozzo\inst{\ref{inst25}a} \and 
M.~del~Prete\inst{\ref{inst59}b} \and 
T.~Dent\inst{\ref{inst54}} \and 
V.~Dergachev\inst{\ref{inst1}} \and 
R.~DeRosa\inst{\ref{inst12}} \and 
R.~DeSalvo\inst{\ref{inst1}} \and 
V.~Dhillon\inst{\ref{inst57}} \and
S.~Dhurandhar\inst{\ref{inst60}} \and 
L.~Di~Fiore\inst{\ref{inst5}a} \and 
A.~Di~Lieto\inst{\ref{inst24}ab} \and 
I.~Di~Palma\inst{\ref{inst7},\ref{inst8}} \and 
M.~Di~Paolo~Emilio\inst{\ref{inst55}ac} \and 
A.~Di~Virgilio\inst{\ref{inst24}a} \and 
M.~D\'iaz\inst{\ref{inst26}} \and 
A.~Dietz\inst{\ref{inst4}} \and
J.~DiGuglielmo\inst{\ref{inst7},\ref{inst8}} \and 
F.~Donovan\inst{\ref{inst21}} \and 
K.~L.~Dooley\inst{\ref{inst11}} \and 
S.~Dorsher\inst{\ref{inst61}} \and 
M.~Drago\inst{\ref{inst59}ab} \and 
R.~W.~P.~Drever\inst{\ref{inst62}} \and 
J.~C.~Driggers\inst{\ref{inst1}} \and 
Z.~Du\inst{\ref{inst45}} \and 
J.-C.~Dumas\inst{\ref{inst20}} \and 
S.~Dwyer\inst{\ref{inst21}} \and 
T.~Eberle\inst{\ref{inst7},\ref{inst8}} \and 
M.~Edgar\inst{\ref{inst3}} \and 
M.~Edwards\inst{\ref{inst54}} \and 
A.~Effler\inst{\ref{inst12}} \and 
P.~Ehrens\inst{\ref{inst1}} \and 
G.~Endr\H{o}czi\inst{\ref{inst58}} \and 
R.~Engel\inst{\ref{inst1}} \and 
T.~Etzel\inst{\ref{inst1}} \and 
K.~Evans\inst{\ref{inst3}} \and 
M.~Evans\inst{\ref{inst21}} \and 
T.~Evans\inst{\ref{inst6}} \and 
M.~Factourovich\inst{\ref{inst23}} \and 
V.~Fafone\inst{\ref{inst55}ab} \and 
S.~Fairhurst\inst{\ref{inst54}} \and 
Y.~Fan\inst{\ref{inst20}} \and 
B.~F.~Farr\inst{\ref{inst63}} \and 
W.~Farr\inst{\ref{inst63}} \and 
D.~Fazi\inst{\ref{inst63}} \and 
H.~Fehrmann\inst{\ref{inst7},\ref{inst8}} \and 
D.~Feldbaum\inst{\ref{inst11}} \and 
I.~Ferrante\inst{\ref{inst24}ab} \and 
F.~Fidecaro\inst{\ref{inst24}ab} \and 
L.~S.~Finn\inst{\ref{inst31}} \and 
I.~Fiori\inst{\ref{inst18}} \and 
R.~P.~Fisher\inst{\ref{inst31}} \and 
R.~Flaminio\inst{\ref{inst33}} \and 
M.~Flanigan\inst{\ref{inst15}} \and 
S.~Foley\inst{\ref{inst21}} \and 
E.~Forsi\inst{\ref{inst6}} \and 
L.~A.~Forte\inst{\ref{inst5}a} \and 
N.~Fotopoulos\inst{\ref{inst1}} \and 
J.-D.~Fournier\inst{\ref{inst32}a} \and 
J.~Franc\inst{\ref{inst33}} \and 
S.~Frasca\inst{\ref{inst14}ab} \and 
F.~Frasconi\inst{\ref{inst24}a} \and 
M.~Frede\inst{\ref{inst7},\ref{inst8}} \and 
M.~Frei\inst{\ref{inst64}} \and 
Z.~Frei\inst{\ref{inst65}} \and 
A.~Freise\inst{\ref{inst13}} \and 
R.~Frey\inst{\ref{inst37}} \and 
T.~T.~Fricke\inst{\ref{inst12}} \and 
J.~K.~Fridriksson\inst{\ref{inst21}} \and
D.~Friedrich\inst{\ref{inst7},\ref{inst8}} \and 
P.~Fritschel\inst{\ref{inst21}} \and 
V.~V.~Frolov\inst{\ref{inst6}} \and 
P.~J.~Fulda\inst{\ref{inst13}} \and 
M.~Fyffe\inst{\ref{inst6}} \and 
M.~Galimberti\inst{\ref{inst33}} \and 
L.~Gammaitoni\inst{\ref{inst35}ab} \and 
M.~R.~Ganija\inst{\ref{inst66}} \and 
J.~Garcia\inst{\ref{inst15}} \and 
J.~A.~Garofoli\inst{\ref{inst19}} \and 
F.~Garufi\inst{\ref{inst5}ab} \and 
M.~E.~G\'asp\'ar\inst{\ref{inst58}} \and 
G.~Gemme\inst{\ref{inst49}} \and 
R.~Geng\inst{\ref{inst45}} \and 
E.~Genin\inst{\ref{inst18}} \and 
A.~Gennai\inst{\ref{inst24}a} \and 
L.~\'A.~Gergely\inst{\ref{inst67}} \and 
S.~Ghosh\inst{\ref{inst34}} \and 
J.~A.~Giaime\inst{\ref{inst12},\ref{inst6}} \and 
S.~Giampanis\inst{\ref{inst9}} \and 
K.~D.~Giardina\inst{\ref{inst6}} \and 
A.~Giazotto\inst{\ref{inst24}a} \and 
C.~Gill\inst{\ref{inst3}} \and 
E.~Goetz\inst{\ref{inst7},\ref{inst8}} \and 
L.~M.~Goggin\inst{\ref{inst9}} \and 
G.~Gonz\'alez\inst{\ref{inst12}} \and 
M.~L.~Gorodetsky\inst{\ref{inst28}} \and 
S.~Go{\ss}ler\inst{\ref{inst7},\ref{inst8}} \and 
R.~Gouaty\inst{\ref{inst4}} \and 
C.~Graef\inst{\ref{inst7},\ref{inst8}} \and 
M.~Granata\inst{\ref{inst22}} \and 
A.~Grant\inst{\ref{inst3}} \and 
S.~Gras\inst{\ref{inst20}} \and 
C.~Gray\inst{\ref{inst15}} \and 
N.~Gray\inst{\ref{inst3}} \and 
R.~J.~S.~Greenhalgh\inst{\ref{inst39}} \and 
A.~M.~Gretarsson\inst{\ref{inst68}} \and 
C.~Greverie\inst{\ref{inst32}a} \and 
R.~Grosso\inst{\ref{inst26}} \and 
H.~Grote\inst{\ref{inst7},\ref{inst8}} \and 
S.~Grunewald\inst{\ref{inst16}} \and 
G.~M.~Guidi\inst{\ref{inst36}ab} \and 
C.~Guido\inst{\ref{inst6}} \and 
R.~Gupta\inst{\ref{inst60}} \and 
E.~K.~Gustafson\inst{\ref{inst1}} \and 
R.~Gustafson\inst{\ref{inst46}} \and 
T.~Ha\inst{\ref{inst69}} \and 
B.~Hage\inst{\ref{inst8},\ref{inst7}} \and 
J.~M.~Hallam\inst{\ref{inst13}} \and 
D.~Hammer\inst{\ref{inst9}} \and 
G.~Hammond\inst{\ref{inst3}} \and 
J.~Hanks\inst{\ref{inst15}} \and 
C.~Hanna\inst{\ref{inst1},\ref{inst70}} \and 
J.~Hanson\inst{\ref{inst6}} \and 
J.~Harms\inst{\ref{inst62}} \and 
G.~M.~Harry\inst{\ref{inst21}} \and 
I.~W.~Harry\inst{\ref{inst54}} \and 
E.~D.~Harstad\inst{\ref{inst37}} \and 
M.~T.~Hartman\inst{\ref{inst11}} \and 
K.~Haughian\inst{\ref{inst3}} \and 
K.~Hayama\inst{\ref{inst71}} \and 
J.-F.~Hayau\inst{\ref{inst32}b} \and 
T.~Hayler\inst{\ref{inst39}} \and 
J.~Heefner\inst{\ref{inst1}} \and 
A.~Heidmann\inst{\ref{inst38}} \and 
M.~C.~Heintze\inst{\ref{inst11}} \and 
H.~Heitmann\inst{\ref{inst32}} \and 
P.~Hello\inst{\ref{inst29}a} \and 
M.~A.~Hendry\inst{\ref{inst3}} \and 
I.~S.~Heng\inst{\ref{inst3}} \and 
A.~W.~Heptonstall\inst{\ref{inst1}} \and 
V.~Herrera\inst{\ref{inst10}} \and 
M.~Hewitson\inst{\ref{inst7},\ref{inst8}} \and 
S.~Hild\inst{\ref{inst3}} \and 
D.~Hoak\inst{\ref{inst42}} \and 
K.~A.~Hodge\inst{\ref{inst1}} \and 
K.~Holt\inst{\ref{inst6}} \and 
J.~Homan\inst{\ref{inst21}} \and
T.~Hong\inst{\ref{inst48}} \and 
S.~Hooper\inst{\ref{inst20}} \and 
D.~J.~Hosken\inst{\ref{inst66}} \and 
J.~Hough\inst{\ref{inst3}} \and 
E.~J.~Howell\inst{\ref{inst20}} \and 
B.~Hughey\inst{\ref{inst9}} \and 
S.~Husa\inst{\ref{inst72}} \and 
S.~H.~Huttner\inst{\ref{inst3}} \and 
T.~Huynh-Dinh\inst{\ref{inst6}} \and 
D.~R.~Ingram\inst{\ref{inst15}} \and 
R.~Inta\inst{\ref{inst52}} \and 
T.~Isogai\inst{\ref{inst51}} \and 
A.~Ivanov\inst{\ref{inst1}} \and 
K.~Izumi\inst{\ref{inst71}} \and 
M.~Jacobson\inst{\ref{inst1}} \and 
H.~Jang\inst{\ref{inst73}} \and 
P.~Jaranowski\inst{\ref{inst40}d} \and 
W.~W.~Johnson\inst{\ref{inst12}} \and 
D.~I.~Jones\inst{\ref{inst74}} \and 
G.~Jones\inst{\ref{inst54}} \and 
R.~Jones\inst{\ref{inst3}} \and 
L.~Ju\inst{\ref{inst20}} \and 
P.~Kalmus\inst{\ref{inst1}} \and 
V.~Kalogera\inst{\ref{inst63}} \and 
I.~Kamaretsos\inst{\ref{inst54}} \and 
S.~Kandhasamy\inst{\ref{inst61}} \and 
G.~Kang\inst{\ref{inst73}} \and 
J.~B.~Kanner\inst{\ref{inst41},\ref{inst30}} \and 
E.~Katsavounidis\inst{\ref{inst21}} \and 
W.~Katzman\inst{\ref{inst6}} \and 
H.~Kaufer\inst{\ref{inst7},\ref{inst8}} \and 
K.~Kawabe\inst{\ref{inst15}} \and 
S.~Kawamura\inst{\ref{inst71}} \and 
F.~Kawazoe\inst{\ref{inst7},\ref{inst8}} \and 
W.~Kells\inst{\ref{inst1}} \and 
D.~G.~Keppel\inst{\ref{inst1}} \and 
Z.~Keresztes\inst{\ref{inst67}} \and 
A.~Khalaidovski\inst{\ref{inst7},\ref{inst8}} \and 
F.~Y.~Khalili\inst{\ref{inst28}} \and 
E.~A.~Khazanov\inst{\ref{inst75}} \and 
B.~Kim\inst{\ref{inst73}} \and 
C.~Kim\inst{\ref{inst76}} \and 
D.~Kim\inst{\ref{inst20}} \and 
H.~Kim\inst{\ref{inst7},\ref{inst8}} \and 
K.~Kim\inst{\ref{inst77}} \and 
N.~Kim\inst{\ref{inst10}} \and 
Y.-M.~Kim\inst{\ref{inst50}} \and 
P.~J.~King\inst{\ref{inst1}} \and 
M.~Kinsey\inst{\ref{inst31}} \and 
D.~L.~Kinzel\inst{\ref{inst6}} \and 
J.~S.~Kissel\inst{\ref{inst21}} \and 
S.~Klimenko\inst{\ref{inst11}} \and 
K.~Kokeyama\inst{\ref{inst13}} \and 
V.~Kondrashov\inst{\ref{inst1}} \and 
R.~Kopparapu\inst{\ref{inst31}} \and 
S.~Koranda\inst{\ref{inst9}} \and 
W.~Z.~Korth\inst{\ref{inst1}} \and 
I.~Kowalska\inst{\ref{inst40}b} \and 
D.~Kozak\inst{\ref{inst1}} \and 
V.~Kringel\inst{\ref{inst7},\ref{inst8}} \and 
S.~Krishnamurthy\inst{\ref{inst63}} \and 
B.~Krishnan\inst{\ref{inst16}} \and 
A.~Kr\'olak\inst{\ref{inst40}ae} \and 
G.~Kuehn\inst{\ref{inst7},\ref{inst8}} \and 
R.~Kumar\inst{\ref{inst3}} \and 
P.~Kwee\inst{\ref{inst8},\ref{inst7}} \and 
M.~Laas-Bourez\inst{\ref{inst20}} \and
P.~K.~Lam\inst{\ref{inst52}} \and 
M.~Landry\inst{\ref{inst15}} \and 
M.~Lang\inst{\ref{inst31}} \and 
B.~Lantz\inst{\ref{inst10}} \and 
N.~Lastzka\inst{\ref{inst7},\ref{inst8}} \and 
C.~Lawrie\inst{\ref{inst3}} \and 
A.~Lazzarini\inst{\ref{inst1}} \and 
P.~Leaci\inst{\ref{inst16}} \and 
C.~H.~Lee\inst{\ref{inst50}} \and 
H.~M.~Lee\inst{\ref{inst78}} \and 
N.~Leindecker\inst{\ref{inst10}} \and 
J.~R.~Leong\inst{\ref{inst7},\ref{inst8}} \and 
I.~Leonor\inst{\ref{inst37}} \and 
N.~Leroy\inst{\ref{inst29}a} \and 
N.~Letendre\inst{\ref{inst4}} \and 
J.~Li\inst{\ref{inst45}} \and 
T.~G.~F.~Li\inst{\ref{inst25}a} \and 
N.~Liguori\inst{\ref{inst59}ab} \and 
P.~E.~Lindquist\inst{\ref{inst1}} \and 
N.~A.~Lockerbie\inst{\ref{inst79}} \and 
D.~Lodhia\inst{\ref{inst13}} \and 
M.~Lorenzini\inst{\ref{inst36}a} \and 
V.~Loriette\inst{\ref{inst29}b} \and 
M.~Lormand\inst{\ref{inst6}} \and 
G.~Losurdo\inst{\ref{inst36}a} \and 
J.~Luan\inst{\ref{inst48}} \and 
M.~Lubinski\inst{\ref{inst15}} \and 
H.~L\"uck\inst{\ref{inst7},\ref{inst8}} \and 
A.~P.~Lundgren\inst{\ref{inst31}} \and 
E.~Macdonald\inst{\ref{inst3}} \and 
B.~Machenschalk\inst{\ref{inst7},\ref{inst8}} \and 
M.~MacInnis\inst{\ref{inst21}} \and 
D.~M.~Macleod\inst{\ref{inst54}} \and 
M.~Mageswaran\inst{\ref{inst1}} \and 
K.~Mailand\inst{\ref{inst1}} \and 
E.~Majorana\inst{\ref{inst14}a} \and 
I.~Maksimovic\inst{\ref{inst29}b} \and 
N.~Man\inst{\ref{inst32}a} \and 
I.~Mandel\inst{\ref{inst21}} \and 
V.~Mandic\inst{\ref{inst61}} \and 
M.~Mantovani\inst{\ref{inst24}ac} \and 
A.~Marandi\inst{\ref{inst10}} \and 
F.~Marchesoni\inst{\ref{inst35}a} \and 
F.~Marion\inst{\ref{inst4}} \and 
S.~M\'arka\inst{\ref{inst23}} \and 
Z.~M\'arka\inst{\ref{inst23}} \and 
A.~Markosyan\inst{\ref{inst10}} \and 
E.~Maros\inst{\ref{inst1}} \and 
J.~Marque\inst{\ref{inst18}} \and 
F.~Martelli\inst{\ref{inst36}ab} \and 
I.~W.~Martin\inst{\ref{inst3}} \and 
R.~M.~Martin\inst{\ref{inst11}} \and 
J.~N.~Marx\inst{\ref{inst1}} \and 
K.~Mason\inst{\ref{inst21}} \and 
A.~Masserot\inst{\ref{inst4}} \and 
F.~Matichard\inst{\ref{inst21}} \and 
L.~Matone\inst{\ref{inst23}} \and 
R.~A.~Matzner\inst{\ref{inst64}} \and 
N.~Mavalvala\inst{\ref{inst21}} \and 
G.~Mazzolo\inst{\ref{inst7},\ref{inst8}} \and 
R.~McCarthy\inst{\ref{inst15}} \and 
D.~E.~McClelland\inst{\ref{inst52}} \and
P.~McDaniel\inst{\ref{inst21}} \and
S.~C.~McGuire\inst{\ref{inst80}} \and 
G.~McIntyre\inst{\ref{inst1}} \and 
J.~McIver\inst{\ref{inst42}} \and
D.~J.~A.~McKechan\inst{\ref{inst54}} \and 
G.~D.~Meadors\inst{\ref{inst46}} \and 
M.~Mehmet\inst{\ref{inst7},\ref{inst8}} \and 
T.~Meier\inst{\ref{inst8},\ref{inst7}} \and 
A.~Melatos\inst{\ref{inst53}} \and 
A.~C.~Melissinos\inst{\ref{inst81}} \and 
G.~Mendell\inst{\ref{inst15}} \and 
D.~Menendez\inst{\ref{inst31}} \and 
R.~A.~Mercer\inst{\ref{inst9}} \and 
S.~Meshkov\inst{\ref{inst1}} \and 
C.~Messenger\inst{\ref{inst54}} \and 
M.~S.~Meyer\inst{\ref{inst6}} \and 
H.~Miao\inst{\ref{inst20}} \and 
C.~Michel\inst{\ref{inst33}} \and 
L.~Milano\inst{\ref{inst5}ab} \and 
J.~Miller\inst{\ref{inst52}} \and 
Y.~Minenkov\inst{\ref{inst55}a} \and 
V.~P.~Mitrofanov\inst{\ref{inst28}} \and 
G.~Mitselmakher\inst{\ref{inst11}} \and 
R.~Mittleman\inst{\ref{inst21}} \and 
O.~Miyakawa\inst{\ref{inst71}} \and 
B.~Moe\inst{\ref{inst9}} \and 
P.~Moesta\inst{\ref{inst16}} \and 
M.~Mohan\inst{\ref{inst18}} \and 
S.~D.~Mohanty\inst{\ref{inst26}} \and 
S.~R.~P.~Mohapatra\inst{\ref{inst42}} \and 
D.~Moraru\inst{\ref{inst15}} \and 
G.~Moreno\inst{\ref{inst15}} \and 
N.~Morgado\inst{\ref{inst33}} \and 
A.~Morgia\inst{\ref{inst55}ab} \and 
T.~Mori\inst{\ref{inst71}} \and 
S.~Mosca\inst{\ref{inst5}ab} \and 
K.~Mossavi\inst{\ref{inst7},\ref{inst8}} \and 
B.~Mours\inst{\ref{inst4}} \and 
C.~M.~Mow--Lowry\inst{\ref{inst52}} \and 
C.~L.~Mueller\inst{\ref{inst11}} \and 
G.~Mueller\inst{\ref{inst11}} \and 
S.~Mukherjee\inst{\ref{inst26}} \and 
A.~Mullavey\inst{\ref{inst52}} \and 
H.~M\"uller-Ebhardt\inst{\ref{inst7},\ref{inst8}} \and 
J.~Munch\inst{\ref{inst66}} \and 
D.~Murphy\inst{\ref{inst23}} \and
P.~G.~Murray\inst{\ref{inst3}} \and 
A.~Mytidis\inst{\ref{inst11}} \and 
T.~Nash\inst{\ref{inst1}} \and 
L.~Naticchioni\inst{\ref{inst14}ab} \and 
R.~Nawrodt\inst{\ref{inst3}} \and 
V.~Necula\inst{\ref{inst11}} \and 
J.~Nelson\inst{\ref{inst3}} \and 
G.~Newton\inst{\ref{inst3}} \and 
A.~Nishizawa\inst{\ref{inst71}} \and 
F.~Nocera\inst{\ref{inst18}} \and 
D.~Nolting\inst{\ref{inst6}} \and 
L.~Nuttall\inst{\ref{inst54}} \and 
E.~Ochsner\inst{\ref{inst41}} \and 
J.~O'Dell\inst{\ref{inst39}} \and 
E.~Oelker\inst{\ref{inst21}} \and 
G.~H.~Ogin\inst{\ref{inst1}} \and 
J.~J.~Oh\inst{\ref{inst69}} \and 
S.~H.~Oh\inst{\ref{inst69}} \and 
R.~G.~Oldenburg\inst{\ref{inst9}} \and 
B.~O'Reilly\inst{\ref{inst6}} \and 
R.~O'Shaughnessy\inst{\ref{inst9}} \and 
C.~Osthelder\inst{\ref{inst1}} \and 
C.~D.~Ott\inst{\ref{inst48}} \and 
D.~J.~Ottaway\inst{\ref{inst66}} \and 
R.~S.~Ottens\inst{\ref{inst11}} \and 
H.~Overmier\inst{\ref{inst6}} \and 
B.~J.~Owen\inst{\ref{inst31}} \and 
A.~Page\inst{\ref{inst13}} \and 
G.~Pagliaroli\inst{\ref{inst55}ac} \and 
L.~Palladino\inst{\ref{inst55}ac} \and 
C.~Palomba\inst{\ref{inst14}a} \and 
Y.~Pan\inst{\ref{inst41}} \and 
C.~Pankow\inst{\ref{inst11}} \and 
F.~Paoletti\inst{\ref{inst24}a,\ref{inst18}} \and 
M.~A.~Papa\inst{\ref{inst16},\ref{inst9}} \and 
M.~Parisi\inst{\ref{inst5}ab} \and 
A.~Pasqualetti\inst{\ref{inst18}} \and 
R.~Passaquieti\inst{\ref{inst24}ab} \and 
D.~Passuello\inst{\ref{inst24}a} \and 
P.~Patel\inst{\ref{inst1}} \and 
M.~Pedraza\inst{\ref{inst1}} \and 
P.~Peiris\inst{\ref{inst82}} \and 
L.~Pekowsky\inst{\ref{inst19}} \and 
S.~Penn\inst{\ref{inst83}} \and 
C.~Peralta\inst{\ref{inst16}} \and 
A.~Perreca\inst{\ref{inst19}} \and 
G.~Persichetti\inst{\ref{inst5}ab} \and 
M.~Phelps\inst{\ref{inst1}} \and 
M.~Pickenpack\inst{\ref{inst7},\ref{inst8}} \and 
F.~Piergiovanni\inst{\ref{inst36}ab} \and 
M.~Pietka\inst{\ref{inst40}d} \and 
L.~Pinard\inst{\ref{inst33}} \and 
I.~M.~Pinto\inst{\ref{inst84}} \and 
M.~Pitkin\inst{\ref{inst3}} \and 
H.~J.~Pletsch\inst{\ref{inst7},\ref{inst8}} \and 
M.~V.~Plissi\inst{\ref{inst3}} \and 
R.~Poggiani\inst{\ref{inst24}ab} \and 
J.~P\"old\inst{\ref{inst7},\ref{inst8}} \and 
F.~Postiglione\inst{\ref{inst56}} \and 
M.~Prato\inst{\ref{inst49}} \and 
V.~Predoi\inst{\ref{inst54}} \and 
L.~R.~Price\inst{\ref{inst1}} \and 
M.~Prijatelj\inst{\ref{inst7},\ref{inst8}} \and 
M.~Principe\inst{\ref{inst84}} \and 
S.~Privitera\inst{\ref{inst1}} \and 
R.~Prix\inst{\ref{inst7},\ref{inst8}} \and 
G.~A.~Prodi\inst{\ref{inst59}ab} \and 
L.~Prokhorov\inst{\ref{inst28}} \and 
O.~Puncken\inst{\ref{inst7},\ref{inst8}} \and 
M.~Punturo\inst{\ref{inst35}a} \and 
P.~Puppo\inst{\ref{inst14}a} \and 
V.~Quetschke\inst{\ref{inst26}} \and 
F.~J.~Raab\inst{\ref{inst15}} \and 
D.~S.~Rabeling\inst{\ref{inst25}ab} \and 
I.~R\'acz\inst{\ref{inst58}} \and 
H.~Radkins\inst{\ref{inst15}} \and 
P.~Raffai\inst{\ref{inst65}} \and 
M.~Rakhmanov\inst{\ref{inst26}} \and 
C.~R.~Ramet\inst{\ref{inst6}} \and 
B.~Rankins\inst{\ref{inst43}} \and 
P.~Rapagnani\inst{\ref{inst14}ab} \and 
S.~Rapoport\inst{\ref{inst52},\ref{addl2}} \and
V.~Raymond\inst{\ref{inst63}} \and 
V.~Re\inst{\ref{inst55}ab} \and 
K.~Redwine\inst{\ref{inst23}} \and 
C.~M.~Reed\inst{\ref{inst15}} \and 
T.~Reed\inst{\ref{inst85}} \and 
T.~Regimbau\inst{\ref{inst32}a} \and 
S.~Reid\inst{\ref{inst3}} \and 
D.~H.~Reitze\inst{\ref{inst11}} \and 
F.~Ricci\inst{\ref{inst14}ab} \and 
R.~Riesen\inst{\ref{inst6}} \and 
K.~Riles\inst{\ref{inst46}} \and 
N.~A.~Robertson\inst{\ref{inst1},\ref{inst3}} \and 
F.~Robinet\inst{\ref{inst29}a} \and 
C.~Robinson\inst{\ref{inst54}} \and 
E.~L.~Robinson\inst{\ref{inst16}} \and 
A.~Rocchi\inst{\ref{inst55}a} \and 
S.~Roddy\inst{\ref{inst6}} \and 
C.~Rodriguez\inst{\ref{inst63}} \and 
M.~Rodruck\inst{\ref{inst15}} \and 
L.~Rolland\inst{\ref{inst4}} \and 
J.~Rollins\inst{\ref{inst23}} \and 
J.~D.~Romano\inst{\ref{inst26}} \and 
R.~Romano\inst{\ref{inst5}ac} \and 
J.~H.~Romie\inst{\ref{inst6}} \and 
D.~Rosi\'nska\inst{\ref{inst40}cf} \and 
C.~R\"{o}ver\inst{\ref{inst7},\ref{inst8}} \and 
S.~Rowan\inst{\ref{inst3}} \and 
A.~R\"udiger\inst{\ref{inst7},\ref{inst8}} \and 
P.~Ruggi\inst{\ref{inst18}} \and 
K.~Ryan\inst{\ref{inst15}} \and 
H.~Ryll\inst{\ref{inst7},\ref{inst8}} \and 
P.~Sainathan\inst{\ref{inst11}} \and 
M.~Sakosky\inst{\ref{inst15}} \and 
F.~Salemi\inst{\ref{inst7},\ref{inst8}} \and 
A.~Samblowski\inst{\ref{inst7},\ref{inst8}} \and
L.~Sammut\inst{\ref{inst53}} \and 
L.~Sancho~de~la~Jordana\inst{\ref{inst72}} \and 
V.~Sandberg\inst{\ref{inst15}} \and 
S.~Sankar\inst{\ref{inst21}} \and 
V.~Sannibale\inst{\ref{inst1}} \and 
L.~Santamar\'ia\inst{\ref{inst1}} \and 
I.~Santiago-Prieto\inst{\ref{inst3}} \and 
G.~Santostasi\inst{\ref{inst86}} \and 
B.~Sassolas\inst{\ref{inst33}} \and 
B.~S.~Sathyaprakash\inst{\ref{inst54}} \and 
S.~Sato\inst{\ref{inst71}} \and 
P.~R.~Saulson\inst{\ref{inst19}} \and 
R.~L.~Savage\inst{\ref{inst15}} \and 
R.~Schilling\inst{\ref{inst7},\ref{inst8}} \and 
S.~Schlamminger\inst{\ref{inst87}} \and 
R.~Schnabel\inst{\ref{inst7},\ref{inst8}} \and 
R.~M.~S.~Schofield\inst{\ref{inst37}} \and 
B.~Schulz\inst{\ref{inst7},\ref{inst8}} \and 
B.~F.~Schutz\inst{\ref{inst16},\ref{inst54}} \and 
P.~Schwinberg\inst{\ref{inst15}} \and 
J.~Scott\inst{\ref{inst3}} \and 
S.~M.~Scott\inst{\ref{inst52}} \and 
A.~C.~Searle\inst{\ref{inst1}} \and 
F.~Seifert\inst{\ref{inst1}} \and 
D.~Sellers\inst{\ref{inst6}} \and 
A.~S.~Sengupta\inst{\ref{inst1}} \and 
D.~Sentenac\inst{\ref{inst18}} \and 
A.~Sergeev\inst{\ref{inst75}} \and 
D.~A.~Shaddock\inst{\ref{inst52}} \and 
M.~Shaltev\inst{\ref{inst7},\ref{inst8}} \and 
B.~Shapiro\inst{\ref{inst21}} \and 
P.~Shawhan\inst{\ref{inst41}} \and 
D.~H.~Shoemaker\inst{\ref{inst21}} \and 
A.~Sibley\inst{\ref{inst6}} \and 
X.~Siemens\inst{\ref{inst9}} \and 
D.~Sigg\inst{\ref{inst15}} \and 
A.~Singer\inst{\ref{inst1}} \and 
L.~Singer\inst{\ref{inst1}} \and 
A.~M.~Sintes\inst{\ref{inst72}} \and 
G.~Skelton\inst{\ref{inst9}} \and 
B.~J.~J.~Slagmolen\inst{\ref{inst52}} \and 
J.~Slutsky\inst{\ref{inst12}} \and 
J.~R.~Smith\inst{\ref{inst2}} \and 
M.~R.~Smith\inst{\ref{inst1}} \and 
N.~D.~Smith\inst{\ref{inst21}} \and 
R.~J.~E.~Smith\inst{\ref{inst13}} \and 
K.~Somiya\inst{\ref{inst48}} \and 
B.~Sorazu\inst{\ref{inst3}} \and 
J.~Soto\inst{\ref{inst21}} \and 
F.~C.~Speirits\inst{\ref{inst3}} \and 
L.~Sperandio\inst{\ref{inst55}ab} \and 
M.~Stefszky\inst{\ref{inst52}} \and 
A.~J.~Stein\inst{\ref{inst21}} \and 
E.~Steinert\inst{\ref{inst15}} \and 
J.~Steinlechner\inst{\ref{inst7},\ref{inst8}} \and 
S.~Steinlechner\inst{\ref{inst7},\ref{inst8}} \and 
S.~Steplewski\inst{\ref{inst34}} \and 
A.~Stochino\inst{\ref{inst1}} \and 
R.~Stone\inst{\ref{inst26}} \and 
K.~A.~Strain\inst{\ref{inst3}} \and 
S.~Strigin\inst{\ref{inst28}} \and 
A.~S.~Stroeer\inst{\ref{inst26}} \and 
R.~Sturani\inst{\ref{inst36}ab} \and 
A.~L.~Stuver\inst{\ref{inst6}} \and 
T.~Z.~Summerscales\inst{\ref{inst88}} \and 
M.~Sung\inst{\ref{inst12}} \and 
S.~Susmithan\inst{\ref{inst20}} \and 
P.~J.~Sutton\inst{\ref{inst54}} \and 
B.~Swinkels\inst{\ref{inst18}} \and 
M.~Tacca\inst{\ref{inst18}} \and 
L.~Taffarello\inst{\ref{inst59}c} \and 
D.~Talukder\inst{\ref{inst34}} \and 
D.~B.~Tanner\inst{\ref{inst11}} \and 
S.~P.~Tarabrin\inst{\ref{inst7},\ref{inst8}} \and 
J.~R.~Taylor\inst{\ref{inst7},\ref{inst8}} \and 
R.~Taylor\inst{\ref{inst1}} \and 
P.~Thomas\inst{\ref{inst15}} \and 
K.~A.~Thorne\inst{\ref{inst6}} \and 
K.~S.~Thorne\inst{\ref{inst48}} \and 
E.~Thrane\inst{\ref{inst61}} \and 
A.~Th\"uring\inst{\ref{inst8},\ref{inst7}} \and 
C.~Titsler\inst{\ref{inst31}} \and 
K.~V.~Tokmakov\inst{\ref{inst79}} \and 
A.~Toncelli\inst{\ref{inst24}ab} \and 
M.~Tonelli\inst{\ref{inst24}ab} \and 
O.~Torre\inst{\ref{inst24}ac} \and 
C.~Torres\inst{\ref{inst6}} \and 
C.~I.~Torrie\inst{\ref{inst1},\ref{inst3}} \and 
E.~Tournefier\inst{\ref{inst4}} \and 
F.~Travasso\inst{\ref{inst35}ab} \and 
G.~Traylor\inst{\ref{inst6}} \and 
M.~Trias\inst{\ref{inst72}} \and 
K.~Tseng\inst{\ref{inst10}} \and 
D.~Ugolini\inst{\ref{inst89}} \and 
K.~Urbanek\inst{\ref{inst10}} \and 
H.~Vahlbruch\inst{\ref{inst8},\ref{inst7}} \and 
G.~Vajente\inst{\ref{inst24}ab} \and 
M.~Vallisneri\inst{\ref{inst48}} \and 
J.~F.~J.~van~den~Brand\inst{\ref{inst25}ab} \and 
C.~Van~Den~Broeck\inst{\ref{inst25}a} \and 
S.~van~der~Putten\inst{\ref{inst25}a} \and 
A.~A.~van~Veggel\inst{\ref{inst3}} \and 
S.~Vass\inst{\ref{inst1}} \and 
M.~Vasuth\inst{\ref{inst58}} \and 
R.~Vaulin\inst{\ref{inst21}} \and 
M.~Vavoulidis\inst{\ref{inst29}a} \and 
A.~Vecchio\inst{\ref{inst13}} \and 
G.~Vedovato\inst{\ref{inst59}c} \and 
J.~Veitch\inst{\ref{inst54}} \and 
P.~J.~Veitch\inst{\ref{inst66}} \and 
C.~Veltkamp\inst{\ref{inst7},\ref{inst8}} \and 
D.~Verkindt\inst{\ref{inst4}} \and 
F.~Vetrano\inst{\ref{inst36}ab} \and 
A.~Vicer\'e\inst{\ref{inst36}ab} \and 
A.~E.~Villar\inst{\ref{inst1}} \and 
J.-Y.~Vinet\inst{\ref{inst32}a} \and 
S.~Vitale\inst{\ref{inst68}} \and 
S.~Vitale\inst{\ref{inst25}a} \and 
H.~Vocca\inst{\ref{inst35}a} \and 
C.~Vorvick\inst{\ref{inst15}} \and 
S.~P.~Vyatchanin\inst{\ref{inst28}} \and 
A.~Wade\inst{\ref{inst52}} \and 
S.~J.~Waldman\inst{\ref{inst21}} \and 
L.~Wallace\inst{\ref{inst1}} \and 
Y.~Wan\inst{\ref{inst45}} \and 
X.~Wang\inst{\ref{inst45}} \and 
Z.~Wang\inst{\ref{inst45}} \and 
A.~Wanner\inst{\ref{inst7},\ref{inst8}} \and 
R.~L.~Ward\inst{\ref{inst22}} \and 
M.~Was\inst{\ref{inst29}a} \and 
P.~Wei\inst{\ref{inst19}} \and 
M.~Weinert\inst{\ref{inst7},\ref{inst8}} \and 
A.~J.~Weinstein\inst{\ref{inst1}} \and 
R.~Weiss\inst{\ref{inst21}} \and 
L.~Wen\inst{\ref{inst48},\ref{inst20}} \and 
S.~Wen\inst{\ref{inst6}} \and 
P.~Wessels\inst{\ref{inst7},\ref{inst8}} \and 
M.~West\inst{\ref{inst19}} \and 
T.~Westphal\inst{\ref{inst7},\ref{inst8}} \and 
K.~Wette\inst{\ref{inst7},\ref{inst8}} \and 
J.~T.~Whelan\inst{\ref{inst82}} \and 
S.~E.~Whitcomb\inst{\ref{inst1},\ref{inst20}} \and 
D.~White\inst{\ref{inst57}} \and 
B.~F.~Whiting\inst{\ref{inst11}} \and 
C.~Wilkinson\inst{\ref{inst15}} \and 
P.~A.~Willems\inst{\ref{inst1}} \and 
H.~R.~Williams\inst{\ref{inst31}} \and 
L.~Williams\inst{\ref{inst11}} \and 
B.~Willke\inst{\ref{inst7},\ref{inst8}} \and 
L.~Winkelmann\inst{\ref{inst7},\ref{inst8}} \and 
W.~Winkler\inst{\ref{inst7},\ref{inst8}} \and 
C.~C.~Wipf\inst{\ref{inst21}} \and 
A.~G.~Wiseman\inst{\ref{inst9}} \and 
H.~Wittel\inst{\ref{inst7},\ref{inst8}} \and 
G.~Woan\inst{\ref{inst3}} \and 
R.~Wooley\inst{\ref{inst6}} \and 
J.~Worden\inst{\ref{inst15}} \and 
J.~Yablon\inst{\ref{inst63}} \and 
I.~Yakushin\inst{\ref{inst6}} \and 
H.~Yamamoto\inst{\ref{inst1}} \and 
K.~Yamamoto\inst{\ref{inst7},\ref{inst8}} \and 
H.~Yang\inst{\ref{inst48}} \and 
D.~Yeaton-Massey\inst{\ref{inst1}} \and 
S.~Yoshida\inst{\ref{inst90}} \and 
P.~Yu\inst{\ref{inst9}} \and 
M.~Yvert\inst{\ref{inst4}} \and 
A.~Zadro\'zny\inst{\ref{inst40}e,\ref{inst91}} \and 
M.~Zanolin\inst{\ref{inst68}} \and 
J.-P.~Zendri\inst{\ref{inst59}c} \and 
F.~Zhang\inst{\ref{inst45}} \and 
L.~Zhang\inst{\ref{inst1}} \and 
W.~Zhang\inst{\ref{inst45}} \and 
Z.~Zhang\inst{\ref{inst20}} \and 
C.~Zhao\inst{\ref{inst20}} \and 
N.~Zotov\inst{\ref{inst85}} \and 
M.~E.~Zucker\inst{\ref{inst21}} \and 
J.~Zweizig\inst{\ref{inst1}} \and
\\
C.~Akerlof\inst{\ref{inst46}} \and
M.~Boer\inst{\ref{addl3}} \and
R.~Fender\inst{\ref{inst74}} \and
N.~Gehrels\inst{\ref{inst30}} \and
A.~Klotz\inst{\ref{addl4}}\and
E.~O.~Ofek\inst{\ref{inst62},\ref{addl5}} \and
M.~Smith\inst{\ref{inst31}} \and
M.~Sokolowski\inst{\ref{inst91}} \and
B.~W.~Stappers\inst{\ref{addl8}} \and
I.~Steele\inst{\ref{addl6}} \and
J.~Swinbank\inst{\ref{addl7}} \and
R.~A.~M.~J.~Wijers\inst{\ref{addl7}} \and
W.~Zheng\inst{\ref{inst46}}
\end{spacing}
}
}

\institute{
\label{inst1}LIGO - California Institute of Technology, Pasadena, CA  91125, USA \and
\label{inst2}California State University Fullerton, Fullerton CA 92831 USA \and
\label{inst3}SUPA, University of Glasgow, Glasgow, G12 8QQ, United Kingdom \and
\label{inst4}Laboratoire d'Annecy-le-Vieux de Physique des Particules (LAPP), Universit\'e de Savoie, CNRS/IN2P3, F-74941 Annecy-Le-Vieux, France \and
\label{inst5}INFN, Sezione di Napoli $^a$; Universit\`a di Napoli 'Federico II'$^b$ Complesso Universitario di Monte S.Angelo, I-80126 Napoli; Universit\`a di Salerno, Fisciano, I-84084 Salerno$^c$, Italy \and
\label{inst6}LIGO - Livingston Observatory, Livingston, LA  70754, USA \and
\label{inst7}Albert-Einstein-Institut, Max-Planck-Institut f\"ur Gravitationsphysik, D-30167 Hannover, Germany \and
\label{inst8}Leibniz Universit\"at Hannover, D-30167 Hannover, Germany \and
\label{inst9}University of Wisconsin--Milwaukee, Milwaukee, WI  53201, USA \and
\label{inst10}Stanford University, Stanford, CA  94305, USA \and
\label{inst11}University of Florida, Gainesville, FL  32611, USA \and
\label{inst12}Louisiana State University, Baton Rouge, LA  70803, USA \and
\label{inst13}University of Birmingham, Birmingham, B15 2TT, United Kingdom \and
\label{inst14}INFN, Sezione di Roma$^a$; Universit\`a 'La Sapienza'$^b$, I-00185 Roma, Italy \and
\label{inst15}LIGO - Hanford Observatory, Richland, WA  99352, USA \and
\label{inst16}Albert-Einstein-Institut, Max-Planck-Institut f\"ur Gravitationsphysik, D-14476 Golm, Germany \and
\label{inst17}Montana State University, Bozeman, MT 59717, USA \and
\label{inst18}European Gravitational Observatory (EGO), I-56021 Cascina (PI), Italy \and
\label{inst19}Syracuse University, Syracuse, NY  13244, USA \and
\label{inst20}University of Western Australia, Crawley, WA 6009, Australia \and
\label{inst21}LIGO - Massachusetts Institute of Technology, Cambridge, MA 02139, USA \and
\label{inst22}Laboratoire AstroParticule et Cosmologie (APC) Universit\'e Paris Diderot, CNRS: IN2P3, CEA: DSM/IRFU, Observatoire de Paris, 10 rue A.Domon et L.Duquet, 75013 Paris - France \and
\label{inst23}Columbia University, New York, NY  10027, USA \and
\label{inst24}INFN, Sezione di Pisa$^a$; Universit\`a di Pisa$^b$; I-56127 Pisa; Universit\`a di Siena, I-53100 Siena$^c$, Italy \and
\label{inst25}Nikhef, Science Park, Amsterdam, the Netherlands$^a$; VU University Amsterdam, De Boelelaan 1081, 1081 HV Amsterdam, the Netherlands$^b$ \and
\label{inst26}The University of Texas at Brownsville and Texas Southmost College, Brownsville, TX  78520, USA \and
\label{inst27}San Jose State University, San Jose, CA 95192, USA \and
\label{inst28}Moscow State University, Moscow, 119992, Russia \and
\label{inst29}LAL, Universit\'e Paris-Sud, IN2P3/CNRS, F-91898 Orsay$^a$; ESPCI, CNRS,  F-75005 Paris$^b$, France \and
\label{inst30}NASA/Goddard Space Flight Center, Greenbelt, MD  20771, USA \and
\label{inst31}The Pennsylvania State University, University Park, PA  16802, USA \and
\label{inst32}Universit\'e Nice-Sophia-Antipolis, CNRS, Observatoire de la C\^ote d'Azur, F-06304 Nice$^a$; Institut de Physique de Rennes, CNRS, Universit\'e de Rennes 1, 35042 Rennes$^b$, France \and
\label{inst33}Laboratoire des Mat\'eriaux Avanc\'es (LMA), IN2P3/CNRS, F-69622 Villeurbanne, Lyon, France \and
\label{inst34}Washington State University, Pullman, WA 99164, USA \and
\label{inst35}INFN, Sezione di Perugia$^a$; Universit\`a di Perugia$^b$, I-06123 Perugia,Italy \and
\label{inst36}INFN, Sezione di Firenze, I-50019 Sesto Fiorentino$^a$; Universit\`a degli Studi di Urbino 'Carlo Bo', I-61029 Urbino$^b$, Italy \and
\label{inst37}University of Oregon, Eugene, OR  97403, USA \and
\label{inst38}Laboratoire Kastler Brossel, ENS, CNRS, UPMC, Universit\'e Pierre et Marie Curie, 4 Place Jussieu, F-75005 Paris, France \and
\label{inst39}Rutherford Appleton Laboratory, HSIC, Chilton, Didcot, Oxon OX11 0QX, United Kingdom \and
\label{inst40}IM-PAN 00-956 Warsaw$^a$; Astronomical Observatory Warsaw University 00-478 Warsaw$^b$; CAMK-PAN 00-716 Warsaw$^c$; Bia{\l}ystok University 15-424 Bia{\l}ystok$^d$; IPJ 05-400 \'Swierk-Otwock$^e$; Institute of Astronomy 65-265 Zielona G\'ora$^f$,  Poland \and
\label{inst41}University of Maryland, College Park, MD 20742 USA \and
\label{inst42}University of Massachusetts - Amherst, Amherst, MA 01003, USA \and
\label{inst43}The University of Mississippi, University, MS 38677, USA \and
\label{inst44}Canadian Institute for Theoretical Astrophysics, University of Toronto, Toronto, Ontario, M5S 3H8, Canada \and
\label{inst45}Tsinghua University, Beijing 100084 China \and
\label{inst46}University of Michigan, Ann Arbor, MI  48109, USA \and
\label{inst47}Charles Sturt University, Wagga Wagga, NSW 2678, Australia \and
\label{inst48}Caltech-CaRT, Pasadena, CA  91125, USA \and
\label{inst49}INFN, Sezione di Genova;  I-16146  Genova, Italy \and
\label{inst50}Pusan National University, Busan 609-735, Korea \and
\label{inst51}Carleton College, Northfield, MN  55057, USA \and
\label{inst52}Australian National University, Canberra, ACT 0200, Australia \and
\label{inst53}The University of Melbourne, Parkville, VIC 3010, Australia \and
\label{inst54}Cardiff University, Cardiff, CF24 3AA, United Kingdom \and
\label{inst55}INFN, Sezione di Roma Tor Vergata$^a$; Universit\`a di Roma Tor Vergata, I-00133 Roma$^b$; Universit\`a dell'Aquila, I-67100 L'Aquila$^c$, Italy \and
\label{inst56}University of Salerno, I-84084 Fisciano (Salerno), Italy and INFN (Sezione di Napoli), Italy \and
\label{inst57}The University of Sheffield, Sheffield S10 2TN, United Kingdom \and
\label{inst58}RMKI, H-1121 Budapest, Konkoly Thege Mikl\'os \'ut 29-33, Hungary \and
\label{inst59}INFN, Gruppo Collegato di Trento$^a$ and Universit\`a di Trento$^b$,  I-38050 Povo, Trento, Italy;   INFN, Sezione di Padova$^c$ and Universit\`a di Padova$^d$, I-35131 Padova, Italy \and
\label{inst60}Inter-University Centre for Astronomy and Astrophysics, Pune - 411007, India \and
\label{inst61}University of Minnesota, Minneapolis, MN 55455, USA \and
\label{inst62}California Institute of Technology, Pasadena, CA  91125, USA \and
\label{inst63}Northwestern University, Evanston, IL  60208, USA \and
\label{inst64}The University of Texas at Austin, Austin, TX 78712, USA \and
\label{inst65}E\"otv\"os Lor\'and University, Budapest, 1117 Hungary \and
\label{inst66}University of Adelaide, Adelaide, SA 5005, Australia \and
\label{inst67}University of Szeged, 6720 Szeged, D\'om t\'er 9, Hungary \and
\label{inst68}Embry-Riddle Aeronautical University, Prescott, AZ   86301 USA \and
\label{inst69}National Institute for Mathematical Sciences, Daejeon 305-390, Korea \and
\label{inst70}Perimeter Institute for Theoretical Physics, Ontario, Canada, N2L 2Y5 \and
\label{inst71}National Astronomical Observatory of Japan, Tokyo  181-8588 \and
\label{inst72}Universitat de les Illes Balears, E-07122 Palma de Mallorca, Spain \and
\label{inst73}Korea Institute of Science and Technology Information, Daejeon 305-806, Korea \and  
\label{inst74}University of Southampton, Southampton, SO17 1BJ, UK \and
\label{inst75}Institute of Applied Physics, Nizhny Novgorod, 603950, Russia \and
\label{inst76}Lund Observatory, Box 43, SE-221 00, Lund, Sweden \and
\label{inst77}Hanyang University, Seoul 133-791, Korea \and
\label{inst78}Seoul National University, Seoul 151-742, Korea \and
\label{inst79}University of Strathclyde, Glasgow, G1 1XQ, United Kingdom \and
\label{inst80}Southern University and A\&M College, Baton Rouge, LA  70813 \and
\label{inst81}University of Rochester, Rochester, NY  14627, USA \and
\label{inst82}Rochester Institute of Technology, Rochester, NY  14623, USA \and
\label{inst83}Hobart and William Smith Colleges, Geneva, NY  14456, USA \and
\label{inst84}University of Sannio at Benevento, I-82100 Benevento, Italy and INFN (Sezione di Napoli), Italy \and
\label{inst85}Louisiana Tech University, Ruston, LA  71272, USA \and
\label{inst86}McNeese State University, Lake Charles, LA 70609 USA \and
\label{inst87}University of Washington, Seattle, WA, 98195-4290, USA \and
\label{inst88}Andrews University, Berrien Springs, MI 49104 USA \and
\label{inst89}Trinity University, San Antonio, TX  78212, USA \and
\label{inst90}Southeastern Louisiana University, Hammond, LA  70402, USA \and
\label{inst91}``Pi of the Sky'' and the Andrzej Soltan Institute for Nuclear Studies, Hoza 69, 00-681 Warsaw, Poland \and
\label{addl2}Research School of Astronomy \& Astrophysics, Mount Stromlo Observatory, Cotter Road, Weston Creek, ACT 2611, Australia \and
\label{addl3}Observatoire de Haute Provence, CNRS, FR 04870 St-Michel l'Observatoire, France \and
\label{addl4}Institut de Recherche en Astrophysique et Planetologie (IRAP), 14 Avenue Edouard Belin, 31400 Toulouse, France \and
\label{addl5}NASA Einstein Fellow \and
\label{addl8}Jodrell Bank Center for Astrophysics, University of Manchester,
Manchester M13 9PL, United Kingdom \and
\label{addl6}Liverpool John Moores University, Liverpool L3 2AJ, United Kingdom \and
\label{addl7}Astronomical Institute ``Anton Pannekoek'', University of Amsterdam, 1090 GE Amsterdam, The Netherlands
}

  \ifthenelse{\equal{\targetjournal}{aa}}{
  }{}
\fi

\ifthenelse{\equal{\targetjournal}{aa}}{

\abstract{} {A transient astrophysical event observed in both 
gravitational wave (GW) and
electromagnetic (EM) channels would yield rich scientific rewards.
A first program initiating EM follow-ups
to possible transient GW events has been developed and exercised by the LIGO and Virgo
community in association with several partners. In this paper, we describe and evaluate the methods used to
promptly identify and localize GW event candidates and to request images
of targeted sky locations.} {During two observing
periods  (Dec 17 2009 to Jan 8 2010 and Sep 2 to Oct 20 2010), a low-latency
analysis pipeline was used to identify GW event candidates and to
reconstruct a map of possible sky locations.  A catalog of nearby galaxies and
Milky Way globular clusters was used to select the most promising
sky positions to be imaged, and this directional
information was delivered to EM observatories with time lags of about thirty
minutes.  A Monte Carlo simulation has been used to evaluate the 
low-latency GW pipeline's ability to  
reconstruct source positions correctly.}
{For signals near the detection threshold,
our low-latency algorithms often localized simulated GW burst signals
to tens of square degrees, while
neutron star/neutron star inspirals and neutron star/black hole
inspirals were localized to a few hundred square degrees.
Localization precision improves for moderately stronger signals.
The correct sky location of
signals well above threshold
and originating from nearby galaxies may be observed with
$\sim$50\% or better probability with a few pointings of wide-field telescopes.}
{}

  \titlerunning{First prompt search for GW transients with EM counterparts}
  \authorrunning{LSC+Virgo+others}
}{
}

\maketitle

 \begin{onecolumn}

\begin{center}
{\bf ABSTRACT}
\end{center}
\vspace{12 pt}
\noindent{\it Aims.} A transient astrophysical event observed in both
gravitational wave (GW) and
electromagnetic (EM) channels would yield rich scientific rewards.
A first program initiating EM follow-ups
to possible transient GW events has been developed and exercised by the LIGO and Virgo
community in association with several partners. In this paper, we describe and evaluate the methods used to
promptly identify and localize GW event candidates and to request images
of targeted sky locations. 

\noindent{\it Methods.}
During two observing
periods  (Dec 17 2009 to Jan 8 2010 and Sep 2 to Oct 20 2010), a low-latency
analysis pipeline was used to identify GW event candidates and to
reconstruct maps of possible sky locations.  A catalog of nearby galaxies and
Milky Way globular clusters was used to select the most promising
sky positions to be imaged, and this directional
information was delivered to EM observatories with time lags of about thirty
minutes.  A Monte Carlo simulation has been used to evaluate the
low-latency GW pipeline's ability to
reconstruct source positions correctly.

\noindent{\it Results.}
For signals near the detection threshold,
our low-latency algorithms often localized simulated GW burst signals
to tens of square degrees, while
neutron star/neutron star inspirals and neutron star/black hole
inspirals were localized to a few hundred square degrees.
Localization precision improves for moderately stronger signals.
The correct sky location of
signals well above threshold
and originating from nearby galaxies may be observed with
$\sim$50\% or better probability with a few pointings of wide-field telescopes.

 \end{onecolumn}

\vspace{12 pt}
\noindent{\bf Key words.} gravitational waves - methods: observational

 \begin{twocolumn}

\section{Introduction}

The Laser Interferometer Gravitational-Wave Observatory (LIGO) \citep{Abbott2009b} and 
Virgo \citep{iVirgo} have taken significant steps toward
gravitational wave
(GW) astronomy over the past decade.
The LIGO Scientific Collaboration operates two LIGO observatories in the U.S. along with
the GEO\,600 detector \citep{GEO-upgrade} in Germany. Together with Virgo,
located in Italy,
they form a detector network capable of detecting GW signals
arriving from all directions.
Their most recent joint data taking run was between July 2009 and October 2010.
GEO\,600 and Virgo are currently operating during summer 2011, while the LIGO 
interferometers have been
decommissioned for the upgrade to the next-generation
Advanced LIGO detectors \citep{advLigo}, expected to be operational around 2015.  
Virgo will also be upgraded to become Advanced Virgo \citep{advVirgo}. Additionally, the new 
LCGT detector \citep{LCGT} is planned in Japan.
These ``advanced era'' detectors are expected to detect compact
binary coalescences, possibly at a rate of dozens per year, after reaching
design sensitivity \citep{cbcRates}.

Detectable, transient GW signals in the LIGO/Virgo frequency band
require bulk motion of mass on short time scales. 
Emission in other channels is also possible in many such rapidly changing 
massive systems.  This leads to the prospect that some transient GW sources
may have corresponding electromagnetic (EM) counterparts which could be discovered
with a low latency response to GW triggers \citep{sylvestre,kanner2008,stubbs2008,kulkarni,bloomDecadal}.  

Finding these EM counterparts would yield rich scientific
rewards (see Sect. \ref{motive}), but is technically 
challenging due to imperfect localization of the gravitational wave signal 
and uncertainty regarding the relative timing of the GW and EM emissions.
This paper details our recent effort
to construct a prompt search for joint GW/EM sources between the LIGO/Virgo
detector network and partner EM observatories (see Sect. \ref{inst}).
The search was demonstrated during two periods of live LIGO/Virgo running:
the ``winter'' observing period in December 2009 and January 2010 and the 
``autumn'' observing period
in September and October 2010.  We focus here on the design and
performance of software developed for rapid EM follow-ups of GW 
candidate events, as well as the procedures used to identify significant
GW triggers and to communicate the most likely sky locations to partner
EM observatories.  
The analysis of the observational data is in progress, and will be the subject 
of future publications.

\section{Motivation} \label{motive}

\subsection{Sources} \label{sources}

A variety of EM emission mechanisms, both observed and theoretical,
may occur in association with observable GW sources.  Characteristics 
of a few scenarios helped inform the design and execution of this search.
Here, some likely models are presented, along with characteristics of the
associated EM emission.

\subsubsection{Compact Binary Coalescence} \label{merger_em}

Compact binary systems consisting of neutron stars and/or black
holes are thought to be the most common sources
of GW emission detectable with ground-based interferometers.
Radiation of energy and angular momentum causes the orbit to
decay (inspiral) until the objects merge \citep{cutlerns}.
For a system consisting 
of two neutron stars (NS-NS) or a neutron star and a 
stellar-mass black hole (NS-BH), the inspiral stage
produces the most readily detectable GW signal.
The energy flux reaching Earth depends on the inclination angle
of the binary orbit relative to the line of sight.
The initial LIGO-Virgo network is sensitive to optimally oriented NS-NS mergers 
from as far away as 30 Mpc, and mergers between a NS and a $10\,M_{\sun}$
black hole out to 70 Mpc \citep{cbcRates}.
Models of the stellar compact object population in the local
universe estimate the rate of NS-NS mergers detectable with initial detectors
to be between $2 \times 10^{-4}$ and $0.2$ per year. 
With advanced detectors, these
range limits are expected to increase to 440 and 930 Mpc, respectively.

The energetics of these systems suggest that an EM counterpart is 
likely.  The final plunge
radiates of order $10^{53}$ ergs of gravitational binding energy
as gravitational waves.
If even a small fraction of this energy escapes
as photons in the observing band, the resulting counterpart could be observable
to large distances.  The EM transients that are likely to follow a NS-BH or 
NS-NS merger are described below.

Short-hard gamma-ray bursts
(SGRBs), which typically have durations of 2 seconds or less, may be powered by NS-NS or 
NS-BH mergers \citep{nakar, meszaros2006, piran2004}.  Afterglows of SGRBs have been observed in wavelengths from
radio to X-ray, and out to Gpc distances \citep{nakar,2009ARA&A..47..567G}.
Optical afterglows have been observed
from a few tens of seconds to a few days after the GRB trigger 
(see, for example, \citet{tarot}), and  
fade with a power law $t^{-\alpha}$, where $\alpha$ is between 1 and 1.5.
At
1 day after the trigger time, the apparent optical magnitude would be between
12 and 20 for a source at 50 Mpc \citep{kannshort}, comparable to the
distance to which LIGO and Virgo could detect the merger.

Even if a compact binary coalescence is not observable in gamma-rays, there is reason
to expect it will produce an observable optical counterpart.
\citet{lipac} suggested that, during a NS-NS or NS-BH merger, some of the neutron star's
mass is tidally ejected.  In their model, the ejected neutron-rich matter
produces heavy elements through {\it r}-process nucleosynthesis,
which subsequently decay and heat the ejecta, powering an optical afterglow
known as a kilonova.
The predicted optical emission is roughly isotropic,
and so is observable regardless of the orientation of the original binary system.
This emission is expected to peak after about one day, around magnitude
18 for a source at 50 Mpc \citep{metzger}, and then fade over the course of a
few days following the merger.

\subsubsection{Stellar Core Collapse}

Beyond the compact object mergers described above, some other astrophysical 
processes are plausible sources of observable
GW emission.  GW transients with unknown waveforms may be discovered by 
searching the LIGO and Virgo data for short periods of excess power (bursts).  
The EM counterparts to some likely sources of GW burst signals are 
described below.  

Core-collapse supernovae are likely to produce some amount of gravitational radiation,
though large uncertainties still exist in the expected waveforms and energetics.
Most models predict GW spectra that would be observable by initial LIGO and Virgo from 
distances within some fraction of the Milky Way, but not from the Mpc distances 
needed to observe GWs from another galaxy \citep{ott}.
Neutrino detectors such as SuperKamiokande and IceCube should also detect a large number of neutrinos from a
Galactic supernova \citep{BeacomVogel,HalzenRaffelt,LeonorEtAl}.
Galactic supernovae normally would be very bright in the optical band,
but could be less than obvious if obscured by dust
or behind the Galactic center.
Optical emission would first appear hours after the GW and neutrino signal
and would peak days to weeks later, fading over the course of weeks or months.

Long-soft gamma-ray bursts (LGRBs) are believed to be associated with the
core collapse of massive stars \citep{woosley93, collapsar1, piran2004, woosley06, metzger11}.  A large variety of possible GW emitting
mechanisms within these systems have been proposed, with some models 
predicting GW spectra that would be observable from distances of
a few Mpc with initial LIGO and Virgo \citep{fryer, kobayashi, corsi, piro07, korobkin11, kiuchi11}.  The afterglows of LGRBs, like the afterglows of 
SGRBs, typically show power law fading with $\alpha = 1 - 1.5$.  
However, the peak isotropic equivalent luminosity of LGRB afterglows 
is typically a factor of 10 brighter than SGRB afterglows \citep{nakar, kann}.

An off-axis or sub-energetic LGRB may also be observed as an orphan afterglow or
dirty fireball \citep{orphans2002, fireballs2003}.  These transients brighten over the course of several
days or even weeks, depending on the observing band and viewing angle.  
Identifying orphan afterglows in large area surveys, such as \citet{rotseorphan}, has proven difficult, but
a GW trigger may help distinguish orphan afterglows from other 
EM variability.   

\subsubsection{Other Possible Sources}

Cosmic string cusps are another possible joint source of GW \citep{cscgw,cscprl} and 
EM \citep{cscem} radiation. 
If present, their distinct GW signature will distinguish them from other sources.
On the other hand, even unmodeled GW emissions can be detected using
GW burst search algorithms, and such events may in some cases produce 
EM radiation either through internal dynamics or through interaction with the 
surrounding medium. Thus, our joint search methods should allow for a wide 
range of possible sources.

\subsection{Investigations enabled by joint GW/EM observations}

A variety of astrophysical
information could potentially be extracted from a joint
GW/EM signal.  In understanding the progenitor physics, 
the EM and GW signals are essentially complementary.
The GW time series directly traces the bulk motion of mass in the 
source,
whereas EM emissions arising from outflows or their interaction with
the interstellar medium give only indirect information requiring
inference and modeling.
On the other hand, observing an EM counterpart to a GW signal
reduces the uncertainty in the source position from degrees to 
arcseconds.  This precise directional information can lead to 
identification of a host galaxy, and a measurement of redshift.
Some specific questions that may be addressed with a collection of 
joint GW/EM signals are discussed below.

If the GW source is identified as a NS-NS or NS-BH merger,
additional investigations are enabled with
an EM counterpart. 
The observation of the EM signal will improve the estimation of astrophysical source parameters.
For example, when attempting parameter estimation
with a bank of templates and a single data stream, the source's
distance, inclination angle, and angular position are largely degenerate.
A precise source position from an EM counterpart would help break this 
degeneracy
\citep{sirens, nissanke}.
High precision parameter estimation may 
even constrain the NS equation of state \citep{cutlerns, vallisneri,
Flanagan2008, andersson, pannarale, hinderer}.  

Observing EM counterparts of NS-NS and NS-BH merger events will give
strong evidence as to which class of source, if either, is the source of SGRBs \citep{bloomDecadal}.  
In addition, if some neutron star mergers are the sources of SGRBs, a collection
of joint EM/GW observations would allow an estimate of the SGRB jet opening angle
by comparing the number of merger events with and without observable prompt EM
emission, and some information would be obtainable even from a single loud
event \citep{Kobayashi2003b, Seto2007}.

An ensemble of these observations could provide a novel measurement
of cosmological parameters.  Analysis of the well-modeled GW signal will provide
a measurement of the luminosity distance to the source, while the redshift distance
is measurable from the EM data.  Taken together, they provide a direct measurement
of the local Hubble constant \citep{hubcon, Markovic1993, sirens, nissanke}.  

Finally, all of the above assume that general relativity is the correct theory
of gravity on macroscopic scales.
Joint EM/GW observations can also be used to test certain predictions of
general relativity, such as the propagation speed and polarizations of GWs \citep{Will2005,
Yunes2010,Kahya2011}.

In the case that the transient GW source is not a binary merger event,
the combination of GW and EM information will again prove very valuable.  
In this scenario, the gravitational waveform will not be known a priori.
Any distance estimate would be derived from the EM data, which would then
set the overall scale for the energy released as GWs.

As in the merger case, the linking of a GW signal with a known 
EM phenomenon will provide insight into the underlying physical process.
For example, the details of the central engine that drives
LGRBs are unknown.  The GW signal could give crucial clues to the motion of 
matter in the source, and potentially distinguish between competing models.
A similar insight into the source mechanism could be achieved for an 
observation of GW emission associated with a supernova.  Rapid identification
may also allow observation of a supernova in its earliest moments,
an opportunity that currently depends on luck \citep{SN2008D}.

\subsection{Extend GW Detector Reach}

Finding an EM counterpart associated with a LIGO/Virgo trigger
would increase confidence that a truly astrophysical event
had been observed in the GW data.  Using EM transients to help distinguish low amplitude
GW signals from noise events allows a lowering of the 
detection threshold, as was done in searches such as \citet{s5grb}.  \citet{kochanek} estimated that 
the detectable amplitude could be reduced by as much as a factor of $1.5$,
increasing the effective 
detector horizon distance (the maximum distance at which an optimally oriented
and located system could be detected) by the same factor 
and thus increasing the detection rate by a factor of 3.
In practice, the actual improvement in GW sensitivity achieved by pairing 
EM and GW observations depends on many factors unique to each search, including details of the 
source model and data set, and so is difficult to predict in advance.

In the case of a coincidence between a GW signal and a
discovered EM transient, the joint significance may be
calculated by assuming that the backgrounds of the EM and
GW search are independent.  The False Alarm Rate (FAR) of
a GW/EM coincidence is the FAR of the GW signal,
as described in Sect. \ref{rankstat}, times $\alpha$,
the expected fraction
of observations associated with a false or unrelated EM
transient.
The false alarm fraction $\alpha$ may be estimated using fields
from surveys not associated with GW triggers.  The measured value of $\alpha$ will
depend heavily on the telescope being used, the cuts selected in image analysis, 
galactic latitude of the source and other factors. 
For example, surveys with the Palomar Transient Factory
require a sophisticated classification mechanism for rejecting
contaminants.  Each set of image subtractions covering 100 - 200 square
degrees yields $\sim 10^5$ candidates.  Of these, 30 - 150 sources
are selected after imposing cuts optimized for the detection
of fast evolving transients \citep{bloomtcp}.  Using
classification software designed for PTF data (Oarical)
\citep{bloomtcp}, the selected sources undergo
an automatic classification as type ``transient''
or ``variable star'' based on time-domain and context
properties.  Promising candidates are selected for additional,
spectroscopic observation.  Of the sources that are classified
as transients, and then followed up spectroscopically, 
$\sim 82\%$ are supernovae \citep{bloomtcp}. 
To use EM transience to improve confidence in a GW signal,
the time-domain sky in the wavelength
of interest must be well understood.  Transients that are found in
directional and time coincidence with GW triggers would increase
confidence in the GW signal only if the chance of a similar, incidental
coincidence is understood to be low \citep{kulkarni}.

\subsection{Implications for Search Design}

Characteristics of the target sources helped determine 
when and where to seek the EM counterparts to GW event candidates.
For reasons discussed in this section, the search strategy presented
in this paper emphasizes
capturing images as soon as possible after the GW trigger, along
with follow-up images over subsequent nights.  The rates of 
stellar core-collapse and compact object mergers
within our own galaxy are much less than one per year, 
and so field selection was 
strongly weighted towards regions containing
nearby galaxies.

The observations and theoretical models of EM transients discussed above
provided guidance when choosing the observing cadence.
GRB optical afterglows have been observed during the prompt emission phase \citep{klotz}
and up to many hours after the trigger. 
For this search, the first attempt to image the source position 
was made as soon as possible after validating a GW trigger.  In both the 
kilonova \citep{lipac} and supernova \citep{ott} models, some time lag exists between the release of GW and
EM emission, primarily due to the time it takes the outflowing material to become
optically thin.  This time lag may be from several hours for a kilonova, up to
days for a core-collapse supernova.  Furthermore, \citet{coward}
show that for GRBs that are off-axis, the optical afterglow may not be visible until days after the burst.
For these reasons, repeated observations over several nights are desirable.   
Also, the light curves obtained by observing the same fields over multiple nights are
critical clues for discovering and classifying transient sources.

Knowing where to look for the counterpart to a GW trigger is challenging.
Directional estimates of low signal to noise ratio (SNR) binary inspiral 
sources with the 2009--10 GW detector network have uncertainties 
of several tens of square degrees \citep{fairhurst}.  This suggests using
telescopes with a field of view (FOV) of at least a few square degrees if 
possible.  Even with such a ``wide field'' instrument, there is a striking 
mismatch between the large area
needing to be searched, and the size of a single FOV.

The problem may be partially mitigated by making use of the known
mass distribution in the nearby universe.  A search for GW counterparts
can dramatically reduce the needed sky coverage by focusing observations 
on galaxies within the distance limits of the GW detectors \citep{kanner2008, nuttall}.  
Limiting the search
area to known galaxies may also improve the feasibility of identifying 
the true counterpart from among other objects with time-varying 
EM emissions \citep{kulkarni}.  
Even within the Milky Way, a search may emphasize known targets by seeking
counterparts within globular clusters, where binary systems of compact objects
may form efficiently \citep{globclust}.
  
An emphasis on extragalactic and globular-cluster sources has the potential drawback that any
counterparts in the plane of the Milky Way may be missed.  Also,
neutron star mergers that occur at large distances from their host galaxies
may not be observed, though the population with large kicks should be small \citep{kicks,zoltan}.  

Our selection of fields to observe was weighted towards areas
containing known galaxies within 50 Mpc.  The utilized catalog of nearby galaxies and
globular clusters, and the process for selecting fields to observe, is described in Sect. 
\ref{tile}.

\section{GW and EM Instruments} \label{inst}

\subsection{Gravitational Wave Detector Network}

The LIGO and Virgo detectors are based on Michelson-type interferometers,
with Fabry-Perot cavities in each arm and a power recycling mirror
between the laser and beamsplitter to dramatically increase the power in
the arms relative to a simple Michelson design.
The GEO\,600 detector uses a folded interferometer without Fabry-Perot
arm cavities but with an additional recycling mirror at the output to
resonantly enhance the GW signal.
As a gravitational wave passes through each interferometer,
it induces a ``strain'' (a minuscule change in
length on the order of 1 part in $10^{21}$ or less) on each arm of the
interferometer due to the quadrupolar perturbation of the spacetime
metric.  The interferometers are designed to measure
the \emph{differential} strain on the two arms through interference of
the laser light when the two beams are recombined at the beam
splitter, with the relative optical phase modulated by the passing
gravitational wave \citep{Abbott2009b}.

In 2009--2010 there were two operating LIGO interferometers, each with
4-km arms: H1, located near Hanford,
Washington, and L1, located in Livingston Parish,
Louisiana.\footnote{Earlier science runs included a second
interferometer at Hanford, called H2, with 2-km arms.  H2 will
reappear as part of Advanced LIGO, either as a second 4-km
interferometer at Hanford or else at a site in Western Australia.
The latter option would greatly improve the source localization
capabilities of the network \citep{Fairhurst2011, schutz2011}.}
Virgo (V1) has arms of
length 3 km and is located near Cascina, Italy.  GEO\,600 data was
not used in the online search described in this paper, but was
available for offline reanalysis of promising event candidates.
The large physical
separation between the instruments means that the effects of local 
environmental background can be mitigated by requiring a coincident signal in
multiple interferometers.  Each interferometer is most sensitive to GW signals traveling
parallel or anti-parallel to zenith, but the antenna pattern varies gradually over
the sky, so that the detectors are essentially all-sky monitors.  

The EM follow-up program described in this paper was exercised
during the 2009--2010 science runs.
While single-detector triggers had been generated with low latency in
earlier science runs for diagnostic and prototyping purposes,
2009--2010 was the first time that a systematic search for GW
transients using the full LIGO-Virgo network was performed with low
latency, and the first time that alerts were sent to external observatories.  

\subsection{Optical and Other Electromagnetic Observatories} \label{scopes}

In an effort to explore various approaches, the telescope network 
used in 2009--10 was intentionally heterogeneous. However, most 
of instruments had large fields of view to accommodate the imprecise GW position
estimate.  The approximate location of each EM
observatory is shown in Fig. \ref{fig:scopemap}, and 
Tables \ref{table:1} and \ref{table:2} show some of the properties of 
each observatory.  

\subsubsection{Optical Instruments}

The \emph{Palomar Transient Factory} (PTF) \citep{PTF2,PTF}
operates a 7.3 square degree FOV camera mounted on the 1.2 m Oschin
Telescope at Palomar Observatory.  A typical
60 s exposure detects objects with a limiting magnitude $R=20.5$. 
For the autumn observing period,
the PTF team devoted ten fields over several nights at
a target rate of 1 trigger for every three weeks. 

\emph{Pi of the Sky} \citep{piofsky} observed using a camera 
with a 400 square degree FOV and exposures to limiting magnitude 11--12. 
It was located in Koczargi Stare, near Warsaw.  The camera was a prototype 
for a planned system that will simultaneously image two steradians of sky.
The target rate was approximately 1 per week in the autumn run,
followed up with hundreds of 10 s exposures over several nights.

The \emph{QUEST} camera \citep{quest}, currently mounted 
on the 1 m ESO Schmidt Telescope at La Silla Observatory,
views 9.4 square degrees of sky
in each exposure. The telescope is capable
of viewing to a limiting magnitude of $R\sim20$.
The QUEST team devoted twelve 60 s exposures
over several nights for each trigger in both
the winter and autumn periods, with a target
rate of 1 trigger per week. 

\emph{ROTSE III} \citep{rotse} is a collection of four robotic
telescopes spread around the world, each with
a 0.45 m aperture and  3.4 square degree FOV.  No filters are used,
so the spectral response is that of the CCDs, spanning roughly 400 to 900 nm.
The equivalent $R$ band limiting magnitude is about 17 in a 20 s exposure.
The
ROTSE team arranged for a series of thirty images for the first night,
and several images on following nights, for each autumn run
trigger, with a target rate of 1 trigger
per week. 

\emph{SkyMapper} \citep{skymap} is a survey telescope located
at Siding Spring observatory in Australia.  The mosaic camera
covers 5.7 square degrees of sky in each field, and is mounted on a 1.35 m 
telescope with a collecting area equivalent to an unobscured 1.01 m aperture.  
It is designed to reach a limiting magnitude $g \sim 21$ ($>$$7$ sigma)
in a 110 s exposure.
SkyMapper
accepted triggers in the autumn run with a target rate of 1 per week,
with several fields collected for each trigger. 

\emph{TAROT} \citep{tarot} operates two robotic 25 cm telescopes, one
at La Silla in Chile and one in Calern, France.  Like the ROTSE III system,
each TAROT instrument has a 3.4 square degree FOV.
A 180 second image with TAROT
in ideal conditions has a limiting $R$ magnitude of $17.5$.  
During the winter run, TAROT observed a single field
during one night for each trigger.  In the autumn run, the field selected
for each trigger was observed over several nights.  TAROT accepted triggers
with a target rate of 1 per week. 

\emph{Zadko Telescope} \citep{zadko} is a 1 m telescope located 
in Western Australia.  The current CCD imager observes fields 
of 0.15 square degrees down to magnitude $\sim20$ in the $R$ band
for a typical 180 s exposure.
For each accepted trigger in the autumn run,
Zadko repeatedly observed the five galaxies considered most likely
to host the source over several nights.  The target
trigger rate for Zadko was one trigger per week. 

The \emph{Liverpool telescope} \citep{liverpool}
is a 2 m robotic telescope situated at the Observatorio
del Roque de Los Muchachos on La Palma. For this project the RATCam instrument,
with a 21 square arcminute field of view, was used. 
This instrumentation allows a five minute exposure to reach magnitude $r'=21$. 
This project was awarded
8 hours of target-of-opportunity time, which was split into 8 observations of 1
hour each, with a target rate of 1 trigger per week. 

\subsubsection{Radio and X-ray Instruments}

\emph{LOFAR} \citep{lofar,2009IEEEP..97.1431D,2011A&A...530A..80S} is a dipole array radio telescope based in
the Netherlands but with stations across Europe.  The array is
sensitive to frequencies in the range of 30 to 80 MHz and 110 to 240
MHz, and can observe multiple simultaneous beams, each with a FWHM
varying with frequency up to a maximum of around 23\textordmasculine.
During the autumn run, LOFAR accepted triggers
at a target rate of 1 per week and followed up each with a four-hour
observation in its higher frequency band, providing
a $\sim$25 square degree field of view.

Although not used in the prompt search during the science run, the
Expanded Very Large Array \citep{evla} was used to follow up a few
triggers after the run with latencies of 3 and 5 weeks.

The \emph{Swift} satellite \citep{swift} carries three instruments,
each in different bands.  Swift granted several
target of opportunity observations with two of these, the X-ray Telescope (XRT)
and UV/Optical Telescope (UVOT), for the winter
and autumn observing periods.  The XRT is an imaging
instrument with a 0.15 square degree FOV, sensitive to fluxes 
around $10^{-13}$ ergs/cm$^2$/s in the 0.5-10 keV band. A few fields were 
imaged for each trigger that Swift accepted.    

\begin{figure}
\begin{center}
\includegraphics[width=1.0\columnwidth]{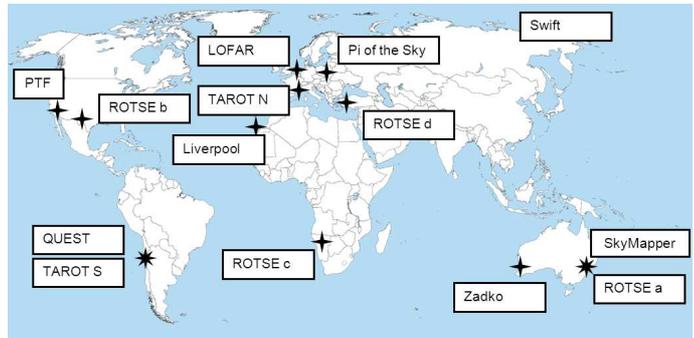}
\caption{A map showing the approximate positions of telescopes
that participated in the project.  The Swift satellite observatory
is noted at an arbitrary location.  The image is adapted from a blank world map
placed in the public domain by P.\ Dlouh{\'y}.} 
\label{fig:scopemap}
\end{center}
\end{figure}

\section{Trigger Selection} \label{TrigSelect}

\begin{figure}
\begin{center}
\includegraphics[width=0.95\columnwidth]{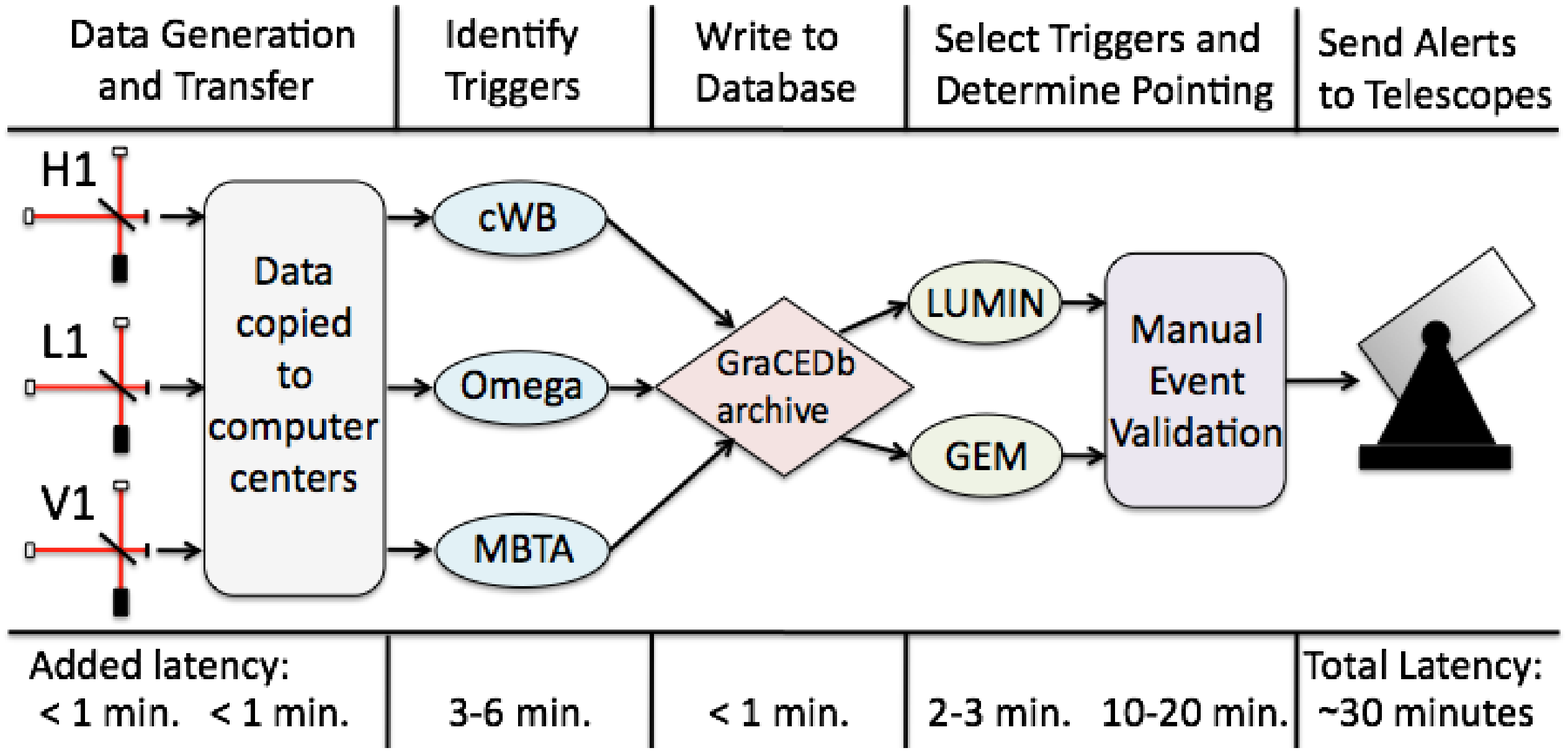}
\caption{A simplified
flowchart of the online analysis with approximate time requirements
for each stage.  Data and information on data quality were generated at the 
Hanford, Livingston, and Virgo
interferometers (H1, L1, and V1) and copied to centralized computer centers.  
The online event trigger generators produced coincident triggers which were 
written into the
GraCEDb archive.  The LUMIN and GEM algorithms selected statistically significant triggers 
from the archive and chose pointing locations.  
Significant triggers generated alerts,
and were validated manually.  If no obvious problem
was found, the trigger's estimated coordinates were sent to telescopes for potential 
follow-up.}
\label{fig:latency}
\end{center}
\end{figure}

The online analysis process which produced GW candidate triggers to be sent to
telescopes is outlined in Fig. \ref{fig:latency}.  After data and 
information on data quality were copied from the interferometer sites to computing centers, three different
data analysis algorithms identified triggers and determined probability skymaps.
The process of downselecting this large collection of triggers
to the few event candidates that received EM follow-up is described in this section.  

After event candidates were placed in
a central archive, additional software used
the locations of nearby galaxies and Milky Way globular clusters
to select likely
source positions (Sect. \ref{tile}).  Triggers were manually vetted, then
the selected targets were passed to partner observatories which 
imaged the sky in an attempt to find an associated
EM transient.  Studies demonstrating the performance
of this pipeline on simulated GWs are presented in Sect. \ref{testing}. 

\subsection{Trigger Generation} \label{triggen}

Sending GW triggers to observatories with less than an hour latency represents
a major shift from past LIGO/Virgo data analyses, which were
reported outside the collaboration at soonest several months after the data 
collection.
Reconstructing source positions requires combining the data streams
from the LIGO-Virgo network using either fully coherent analysis or a
coincidence analysis of single-detector trigger times.
A key step in latency reduction was the rapid data replication process,  
in which data from all three GW observatory sites were copied to several 
computing centers within a minute of collection.

For the EM follow-up program, three independent GW detection algorithms (trigger generators),
ran promptly as data became available, generating candidate triggers with latencies
between five and eight minutes. Omega Pipeline and 
coherent WaveBurst (cWB), which are both described in \citet{Abadie2010},
searched for transients (bursts) with only loose assumptions regarding waveform morphology. 
The Multi-Band Template Analysis (MBTA) \citep{mbta}, searched for 
signals from coalescing compact binaries.
Triggers
were ranked by their ``detection statistic'', a figure of merit for
each analysis, known as $\Omega$, $\eta$, and $\rho_{\rm{combined}}$, respectively. 
The statistics $\eta$ for cWB and $\rho_{\rm{combined}}$ for MBTA are 
related to the amplitude SNR of the signal across all interferometers while $\Omega$ is
related to the Bayesian likelihood of a GW signal being present.
Triggers with a detection statistic above a nominal
threshold, and occurring in times where all three detectors were operating normally,
were recorded in the Gravitational-wave Candidate Event Database (GraCEDb).

The trigger generators also produced likelihood maps over the sky (skymaps), indicating
the location from which the signal was most likely to have originated.  A brief introduction
to each trigger generator is presented 
in Sects. \ref{trig_cwb} -- \ref{trig_mbta}. 

\subsubsection{Coherent WaveBurst} \label{trig_cwb}

Coherent WaveBurst has been used in previous searches for GW bursts,
such as \citet{Abbott2009} and \citet{Abadie2010}.
The algorithm performs a time-frequency analysis of data in the wavelet domain.
It coherently combines data from all detectors 
to reconstruct the two GW polarization waveforms 
$h_+(t)$ and $h_\times(t)$ and the source coordinates on the sky. 
A statistic is constructed from the coherent terms of the maximum likelihood
ratio functional~\citep{FH1998,PRD05} for each possible sky location,
and is used to rank each location in a grid that 
covers the sky (skymap). 
A detailed description of the likelihood analysis, the sky localization
statistic and the performance of the cWB algorithm is published 
elsewhere~\citep{cwbposrec}.

The search was run in two configurations which differ
in their assumptions about the GW signal.  The ``unconstrained'' search
places minimal assumptions on the GW waveform, while the ``linear'' search
assumes the signal is dominated by a single GW polarization state
\citep{cwbposrec}.  
While the unconstrained search is more general, 
and is the configuration that was used in previous burst analyses,
the linear search has been
shown to better estimate source positions for some classes of signals.
For the online analysis, the two searches were run in parallel.

\subsubsection{Omega Pipeline} \label{trig_omg}

In the Omega Pipeline search \citep{Abadie2010}, triggers are first identified by performing a matched
filter search with a bank of basis waveforms which are approximately 
(co)sine-Gaussians. The search assumes that a GW signal can be decomposed
into a small number of these basis waveforms,  and so is most sensitive to 
signals with a small time-frequency volume.  Coincidence criteria are then
applied, requiring a trigger with consistent frequency in another 
interferometer within a physically consistent time window. A coherent 
Bayesian position reconstruction code \citep{searle2008,searle2009} is then applied to remaining candidates.
 The code performs Bayesian marginalization over all parameters 
(time of arrival, amplitude and polarization) other than direction. 
This results in a skymap providing the probability that a signal arrived at 
any time, with any amplitude and polarization, as a function of direction.  
Further 
marginalization is performed over this entire probability skymap to arrive at
a single number, the estimated probability that a signal arrived from any direction. 
The $\Omega$ statistic is constructed from this number and other trigger 
properties.

\subsubsection{MBTA} \label{trig_mbta}

The Multi-Band Template Analysis (MBTA) is a low-latency implementation of the
matched filter search that is typically used to search for compact binary
inspirals \citep{mbta, cbcquick}.  In contrast to burst searches which do
not assume any particular waveform morphology, MBTA specifically
targets the waveforms expected from NS-NS, NS-BH and BH-BH
inspirals.  In this way it provides complementary coverage to the burst
searches described above.

The 
search uses templates computed from a second order post-Newtonian 
approximation for the phase evolution of the signal, with component 
masses in the range $1$--$34\,M_{\sun}$ and a total mass of $< 35\,M_{\sun}$.  
However, triggers generated from
templates with {\em both} component masses larger than the plausible
limit of the NS mass---conservatively taken to be $3.5\,M_{\sun}$ for
this check---were not considered for EM follow-up, since the optical
emission is thought to be associated with the merger of two neutron stars or with the disruption of a neutron star by a stellar-mass black hole.

Triggers from each interferometer
are clustered and used to search for coincidence among the individual 
detectors.  To generate a candidate event for follow-up, triggers with 
consistent physical parameters must be present in all
three LIGO/Virgo interferometers.
For each triple 
coincident trigger, the sky location was estimated using 
the time delay between detector sites and the amplitude of the signal measured 
in each detector \citep{fairhurst}.
Before the observing period, a set of simulated gravitational wave signals was
used to measure the distribution of errors in recovering the time delays and 
signal amplitudes.  The sky localization algorithm then uses these distributions
to assign probabilities to each pixel on the sky.

\subsection{Estimating False Alarm Rates} \label{rankstat}

The primary quantity used to decide whether an event should be considered a
candidate for follow-ups was its FAR, the average rate at which
noise fluctuations create events
with an equal or greater value of the detection statistic.
For the winter run, a FAR of less than 1 event per day of livetime was
required to send an imaging request to the ground-based telescopes,
with a higher threshold for Swift.
For the autumn run, the FAR threshold was 0.25 events per day of livetime
for most
telescopes, with stricter requirements for sending triggers to Palomar Transient
Factory and Swift.
Livetime is here defined as time all three interferometers
were simultaneously collecting usable science data.

As in previous all-sky burst searches, e.g. \citet{Abbott2009}
and \citet{Abadie2010}, 
the FAR for the two burst pipelines was evaluated using the time-shift method.
In this method, artificial time shifts, between one second and a few hundred seconds,
are applied to the strain series of one or
more interferometers, and the shifted data streams are analyzed with the
regular coherent search algorithm.  The shifted data provide an 
estimate of the background noise trigger rate without any true coincident 
gravitational wave signals.  During the online analysis, at least 100 
time shifts were continuously evaluated with latencies between 10 minutes and 
several hours.  The FAR of each event candidate was evaluated with the most recent
available time shifts.

The MBTA pipeline evaluated the FAR analytically based on single interferometer
trigger rates, rather than using time shifts.  This is computationally simpler 
than the burst method.  It is valid since MBTA is a coincident rather than a 
coherent analysis, and allows the FAR to be evaluated with data from the 
minutes immediately preceding the trigger time \citep{mbta}.

\subsection{Online Data Quality} 

A number of common occurrences may make a stretch of interferometer data
unsuitable for sensitive GW searches.  Examples include times of 
large seismic disturbance, non-standard interferometer configurations,
and temporary saturations of various photodiodes in the interferometer
sensing and control system.
To mark such times, monitor programs analyze auxiliary data
to produce lists of abnormal time segments with low latency. 
When a trigger was identified, 
it was automatically checked against these 
lists; triggers which occurred in stretches of unacceptable 
data were automatically rejected.  During this search,
all three GW detectors were simultaneously collecting
science quality data for roughly $45\%$ of the time. 

\subsection{Manual Event Validation} \label{human}

In addition to automated checks on data quality, significant triggers
were manually vetted.  Trigger alerts were broadcast to collaboration 
members via e-mail, text message, a website, and in the interferometer
control rooms as audio alarms.  For each alert, a
low-latency pipeline expert conferred with personnel at each
of the three observatory sites to
validate the event.  Pipeline experts and scientists monitoring data on-site  
provided 24/7 coverage in 8 hour shifts.
Assigned personnel confirmed
the automated data quality results, checked plots for obvious
abnormalities, and verified that there were no known
disturbances at any of the three observatory sites.

The intention of manual event validation was to veto
spurious events
caused by known non-GW mechanisms that have not been caught by low-latency
data quality cuts, not to remove every non-GW 
trigger.  In fact, at current sensitivities, most or all of the triggers are 
unlikely to represent true astrophysical events.   The trade-off
for this additional check on the quality of the triggers
was added latency (usually 10 to 20 minutes) between trigger identification
and reporting to the EM observatories.  It is possible
that as the search
matures in the Advanced LIGO/Virgo era the validation process can be fully
automated.

\section{Choosing Fields to Observe} \label{tile}

The uncertainty associated with GW position estimates,
expected to be several tens of square degrees \citep{fairhurst}, is 
large compared to the FOV of most
astronomical instruments.
Moreover, the likely sky regions
calculated from interferometer data may be irregularly shaped, or
even contain several disjoint regions.  It is impractical to image these
entire regions given a limited amount of observing time for a given instrument.
There is thus a need to carefully prioritize fields, or tiles,
of an instrument to optimize the likelihood of imaging the true
gravitational wave source. 

The LUMIN software package was created to gather GW triggers from the 
three trigger generators, and use the skymaps and locations of known galaxies to select
fields for each optical or radio instrument to observe.  In addition, LUMIN
includes tools that were used to facilitate trigger validation (Sect. \ref{human})
and communication with robotic telescopes.  Fields for observation with the 
Swift XRT and UVOT were selected with slightly different criteria by a separate
software package, the Gravitational to Electro-Magnetic Processor (GEM).  During the testing 
process, GEM also applied the tiling criteria for optical telescopes to 
simulated skymaps, and so provided an important consistency check between LUMIN and GEM.

\subsection{Galaxy Catalog}
\label{Galaxy Catalog}
The Gravitational Wave Galaxy Catalog (GWGC) \citep{GWGC} was created  
to help this and future searches quickly identify nearby galaxies.

The catalog contains up-to-date information compiled from the
literature on sky position, distance, blue magnitude, major and minor
diameters, position angle and galaxy type for 53,225 galaxies ranging
out to 100 Mpc, as well as 150 Milky Way globular clusters.
\citet{GWGC} compared the catalog with an expected blue light
distribution derived from SDSS data and concluded that the GWGC is
nearly complete out to $\sim$40 Mpc.
The catalog improves on the issue of multiple entries for the same galaxy 
suffered by previous catalogs by creating the GWGC from a subset of 4 large 
catalogs, each of which lists a unique Principal Galaxy Catalogue (PGC) number 
for every galaxy \citep{pgc}. 
The catalogs used were: an updated version of the Tully Nearby Galaxies Catalog \citep{tully3k},
the Catalog of Neighboring Galaxies \citep{cng}, the V8k catalog \citep{edd}, and HyperLEDA \citep{hyperleda}.
Also included is a list of 150 
known Milky Way globular clusters \citep{mwgc}. These 
are all available freely online, 
but a local, homogeneous list is essential for rapid follow-up purposes.

\subsection{Weighting and Tiling Algorithm}
To make use of the galaxy catalog, and choose tiles for 
each GW trigger, similar algorithms have been implemented in
the GEM and LUMIN software packages.  

The position information from the trigger generators 
(see Sect. \ref{triggen}) is encoded in skymaps that
assign a likelihood to each $0.4^{\circ}\times0.4^{\circ}$ pixel in a grid covering the sky.
In practice, only the 1000 most likely pixels are retained, limiting
the sky area to roughly 160 square degrees.  
The search volume is further limited by keeping
only objects in the catalog with an estimated distance of less than 50 Mpc,
as the current sensitivity of the GW detectors makes it unlikely that binaries
containing a neutron star would be detectable beyond this distance.  
Approximately 8\% of the pixels in an
average skymap contain a local galaxy or globular cluster listed in the GWGC catalog.

For burst triggers, the tiling algorithms treat the luminosity of each galaxy or globular cluster 
as a prior for its likelihood to host a GW emitting event. 
The blue light luminosity is used as a proxy for star formation, indicating
the presence of massive stars that may be GW burst progenitors themselves and
may evolve into compact binaries that eventually merge.
In addition, 
weak sources of GWs are assumed to be more numerous than strong sources, so that a closer
galaxy should contain more {\em detectable} sources than a more distant
galaxy of the same mass \citep{nuttall}.
This leads to assigning the following likelihood to each pixel:
\begin{equation}
P \propto \displaystyle\sum\limits_{i} \frac{M_i L}{D_i}
\label{eqn:galweight}
\end{equation}
where $L$ is the likelihood based only on the GW data, and $M$ and $D$
are the blue light luminosity (a rough proxy for mass) and distance of the 
associated galaxy or globular cluster.  
The sum is over all the objects 
associated with a particular pixel (which will be 0 or 1 galaxy for the 
majority of pixels).  Extended nearby sources which have a major axis larger than the pixel size
have their mass divided evenly over each pixel falling within the ellipse of the disk defined by their major and minor axes.  Once this calculation is 
performed for each pixel, the entire skymap is renormalized to a total 
likelihood equal to unity.

Unlike the burst algorithms, MBTA assumes the GW source is a merging binary, 
and estimates some of the source's physical parameters for each
trigger.  This allows the galaxy catalog to be applied in a slightly different way.  Each
interferometer measures a quantity known as {\it effective distance}
\begin{equation}
D_{\mathrm {eff}} = D\left[ F_{+}^2\left(\frac{1+\cos^2\iota}{2}\right)^2 +
F_{\times}^2\cos^2\iota\right]^{-1/2},\label{Deff}
\end{equation}
where $D$ is the actual distance to the source, $\iota$ is the inclination angle between the
direction to the observer and the angular momentum vector of the binary, and
$F_{+}$ and $F_{\times}$ are the antenna response functions of the particular
interferometer.  The important feature of the effective distance is that
it is always greater than or equal to the true distance to the source.  For each
MBTA trigger the galaxy catalog is then only considered out to the smallest
effective distance measured for that trigger, with a maximum possible
effective distance of
50 Mpc.  After the 
catalog is downselected in this way, each pixel is weighted by the fraction of
the catalog's total mass contained in that pixel, i.e.
\begin{equation}
P = \displaystyle\sum\limits_{i} M^{\mathrm {frac}}_i L,
\end{equation}
with the sum over all galaxies associated with the pixel, and $\sum_k M^{\mathrm {frac}}_k = 1$
for a sum over the downselected catalog.

These procedures require a pixel's coordinates to be consistent with a 
known galaxy's location
to be targeted by telescopes.  
However, in the case that the skymap does not 
intersect with any galaxies in the catalog, the likelihood from the GW skymap
alone is used as each pixel's likelihood ($P=L$).  In practice, this is a very
rare occurrence and only happens in the case of a very well-localized skymap.

The actual pointing coordinates requested for each telescope are 
selected to maximize the total contained $P$ summed over pixels
within the FOV.  If multiple pointings are allowed with
the same instrument, additional tiles with the next highest ranking are
then selected.  The tile selection process is illustrated in Fig. \ref{skymap}.

\ifthenelse{\equal{\targetjournal}{aa}}{\begin{figure*}}{\begin{figure*}}
\begin{center}
\mbox{
\includegraphics[width=0.45\textwidth]{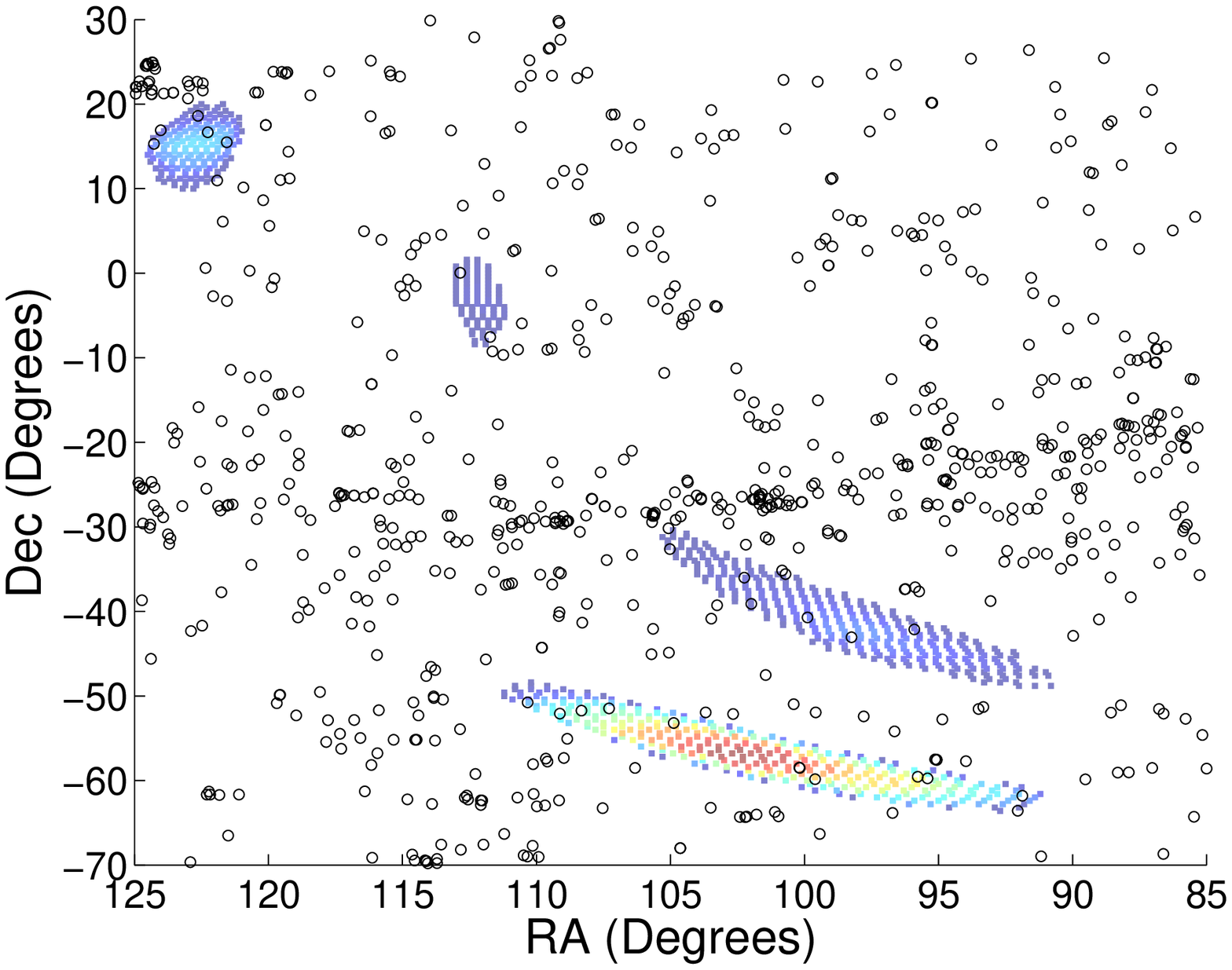} 
\hspace{0.5cm}
\includegraphics[width=0.45\textwidth]{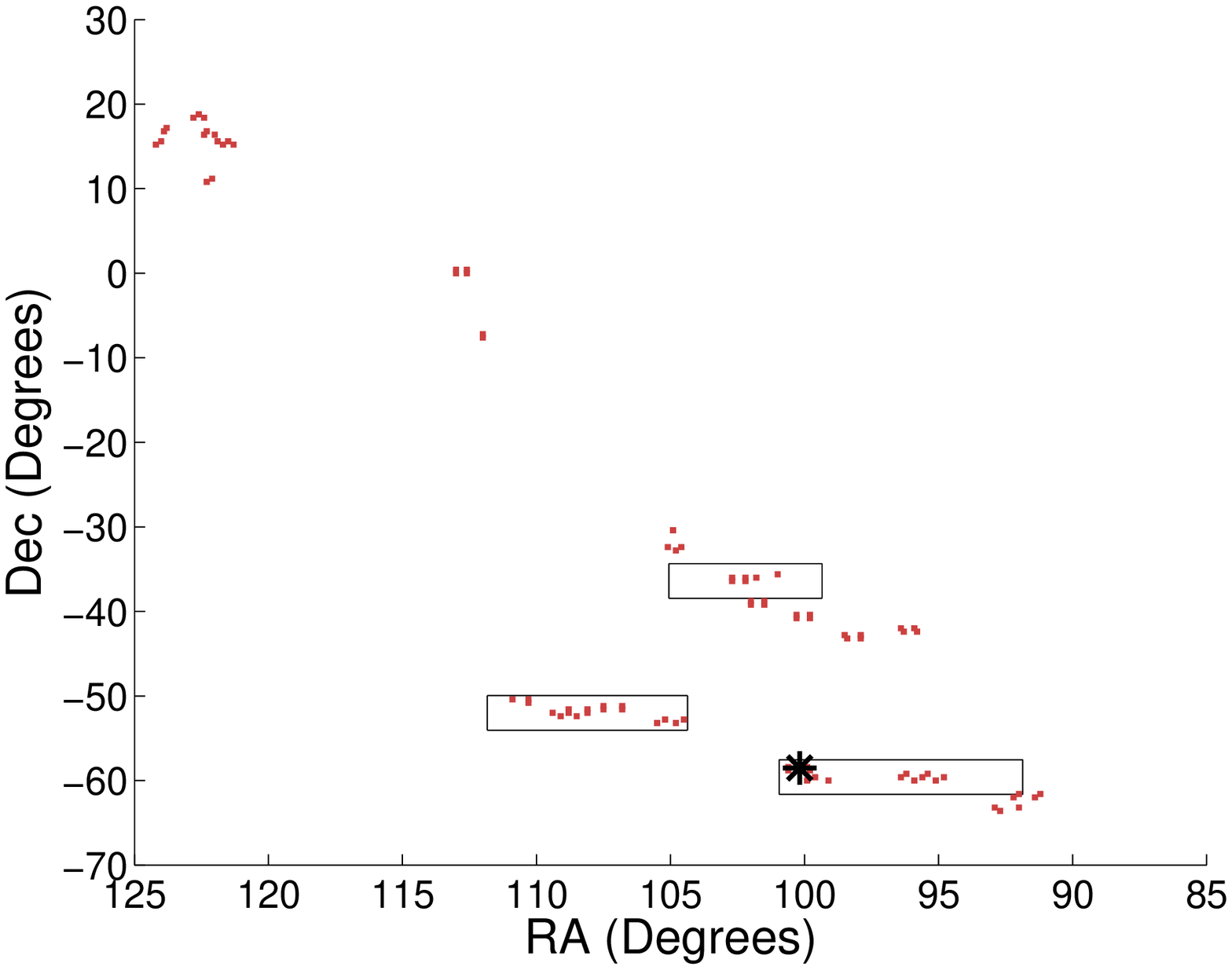}
}

\caption{The weighting and tiling process for a simulated signal
reconstructed by cWB.  The skymap is shown in the left panel with the highest 
likelihood regions in red,
and lower ranked pixels in blue,
along with galaxy locations marked as black circles.  
The right panel shows the location and approximate size of the three chosen QUEST tiles,
along with the locations of pixels that are retained after weighting by the galaxy catalog.
The injection location is caught by the southernmost tile, and is marked with an asterisk.} 
\label{skymap}
\end{center}
\ifthenelse{\equal{\targetjournal}{aa}}{\end{figure*}}{\end{figure*}}

\subsection{Galaxy Targeting for Small-Field Instruments}

The logic used for selecting pointings for the Swift satellite was similar to
that of ground-based telescopes, except that, because the narrower Swift FOV
required greater precision, care was taken to ensure
the target galaxies were within the selected field.  
The coordinates supplied to Swift for 
follow-up were those of the matched galaxy itself in cases where there was
only a single galaxy in a pixel, but the center of the 
0.4$^{\circ}\times$0.4$^{\circ}$ pixel in cases where the central coordinates 
of an extended source were outside the pixel or there were multiple galaxies
in the pixel.  Since fewer follow-ups were allowed using
Swift than with other scopes, a minimum requirement was placed on the
statistic $P$ contained within the pixels selected for X-ray observation.

Zadko and Liverpool Telescope also have relatively narrow fields.  For these
telescopes, no attempt was made to capture multiple galaxies in a single field.  
Instead, the weighting scheme in Eqn. \ref{eqn:galweight} was applied to each
galaxy rather than each pixel, and the center coordinates of the top ranked
galaxies were passed to the observatories.

\section{Observing Strategy} \label{strategy}

\subsection{Communication}

After an event candidate passed manual inspection,
a script was launched to pass the GPS time and 
selected field center locations to the QUEST, ROTSE III, 
SkyMapper, TAROT, Zadko, Liverpool Telescope, and LOFAR 
observatories.  During the autumn run, a total of five
such alerts were sent.  During the (earlier)
winter run, 8 event candidates were passed to the TAROT
and QUEST observatories.
The number of field locations passed to each telescope 
for each GW event candidate are listed as the ``Tiles per Trigger'' 
in Table \ref{table:2}.  During the autumn run, 
in cases where the fields selected for 
a particular instrument were unobservable due to 
daylight or latitude, no alert was sent to 
the observatory.
In most cases, alerts were sent via a 
direct socket connection from a LIGO 
computing center at Caltech with IP mask
protection.  Alerts
to ROTSE III, SkyMapper, TAROT, and Zadko
used the format of GCN notices.  Alerts to 
LOFAR and the Liverpool Telescope
used the VOEvent format \citep{2006ASPC..351..637W}.  
For QUEST, the GPS time
and field positions were posted as 
ASCII tables to a password protected web
site which was regularly polled by the
QUEST scheduler.  

The Palomar Transient Factory received
field locations and GPS times using the 
VOEvent format via a socket connection, 
but with a more restrictive FAR threshold 
than the other optical telescopes, 
and so triggers were only sent to PTF
if the on-call team executed a separate 
script.  Alerts to Swift also required 
extra action by the on-call team, who
entered field coordinates in an online form.
The Pi of the Sky prototype telescope was engaged 
through automated e-mails and manual checks of a 
password protected web page. 

\subsection{Telescope Response}

The wide variety of telescopes involved in the 
search led to a diversity of observing strategies,
with each partnering group applying a different cadence.
By design, most of the telescopes in the 
network were robotic, and could respond to alerts
without human intervention.  In a few cases this allowed 
response times of less than a minute
after an alert was sent, though response times of 
a few hours were more typical due to wait time
for targets to be overhead.

During the winter run, QUEST responded to three
triggers, making 2 exposures of each field 
on the night of the request.  TAROT responded
to one winter run trigger, taking six images 
on the night of the request.  Swift also 
responded to one trigger in the winter 
run, taking one exposure of each field 
following the request, and then a second
set of exposures on a later date to be used
as reference images.  

For most observatories in the summer run, 
the observing plan called for capturing a first image
of the selected fields as rapidly
as possible, with follow-up observations
every night or every other night out to 
five days after the trigger time.  
For the optical observatories, any night's
observation included 2 or more exposures
for each field, to help eliminate asteroids,
CCD artifacts, and other contaminants
from the data set.  In addition, some
fields were imaged at later times, up to 
a month after the trigger time, to 
provide reference images, or possibly to 
capture a light curve with a late brightening time.
TAROT, Zadko, PTF, QUEST, and Pi of the Sky
all followed this recipe.  
ROTSE executed a more aggressive observing plan, 
collecting a set of 30 images in rapid succession on 
the first night, and then sets of eight images 
on each of 15 nights following the trigger with 
intervals of two days on average.
As in the winter run, Swift made one
exposure of each field following the trigger, and then
collected a reference image after a lag of several
weeks.  The Liverpool Telescope devoted 
roughly one hour of observation upon receiving a 
trigger, and then collected reference images a few
weeks after the trigger time.  
The LOFAR response was not automated.  A telescope
operator made a single, four hour observation one to 
four days after delivery of a trigger.  SkyMapper also 
required manual intervention to respond to a trigger,
and so responded on a best effort basis.

\section{Performance Study}\label{testing}

\subsection{Simulated Waveform Injections} \label{etgTesting}

\ifthenelse{\equal{\targetjournal}{aa}}{\begin{figure*}}{\begin{figure*}}
\begin{center}
\mbox{
\includegraphics*[width=0.38\textwidth]{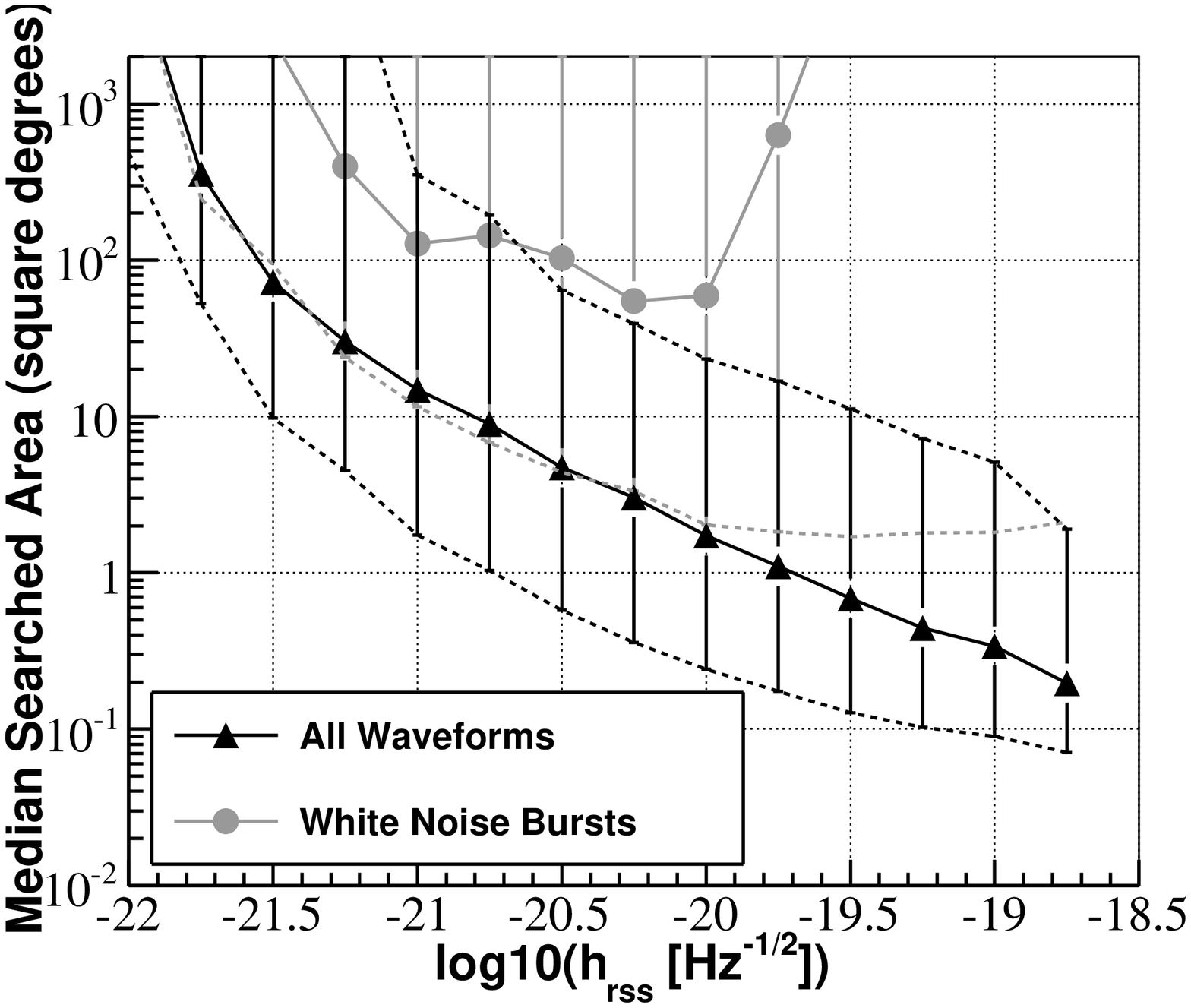} 
\hspace{0.5cm}
\includegraphics*[width=0.38\textwidth]{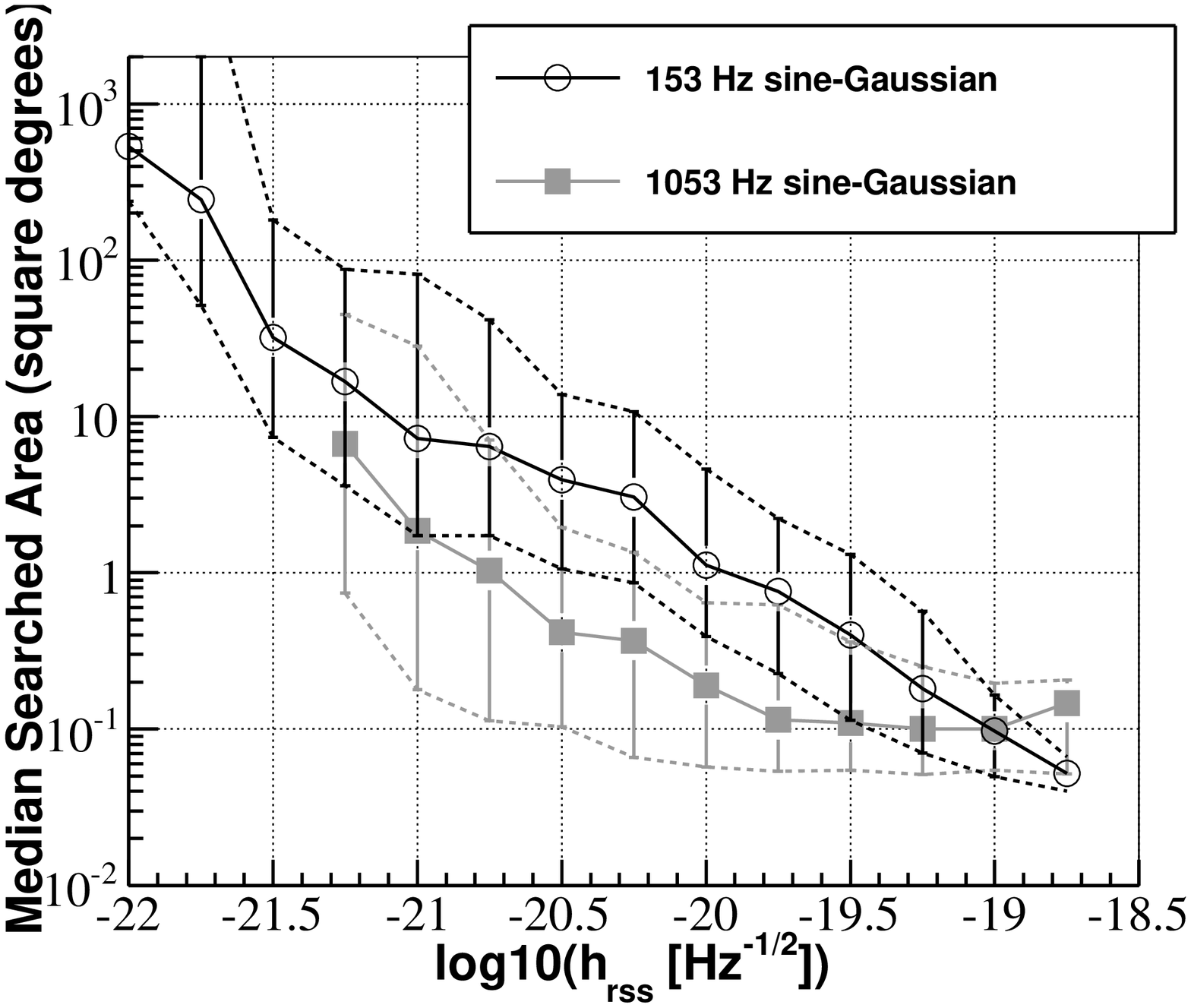}
}
\mbox{
\includegraphics*[width=0.38\textwidth]{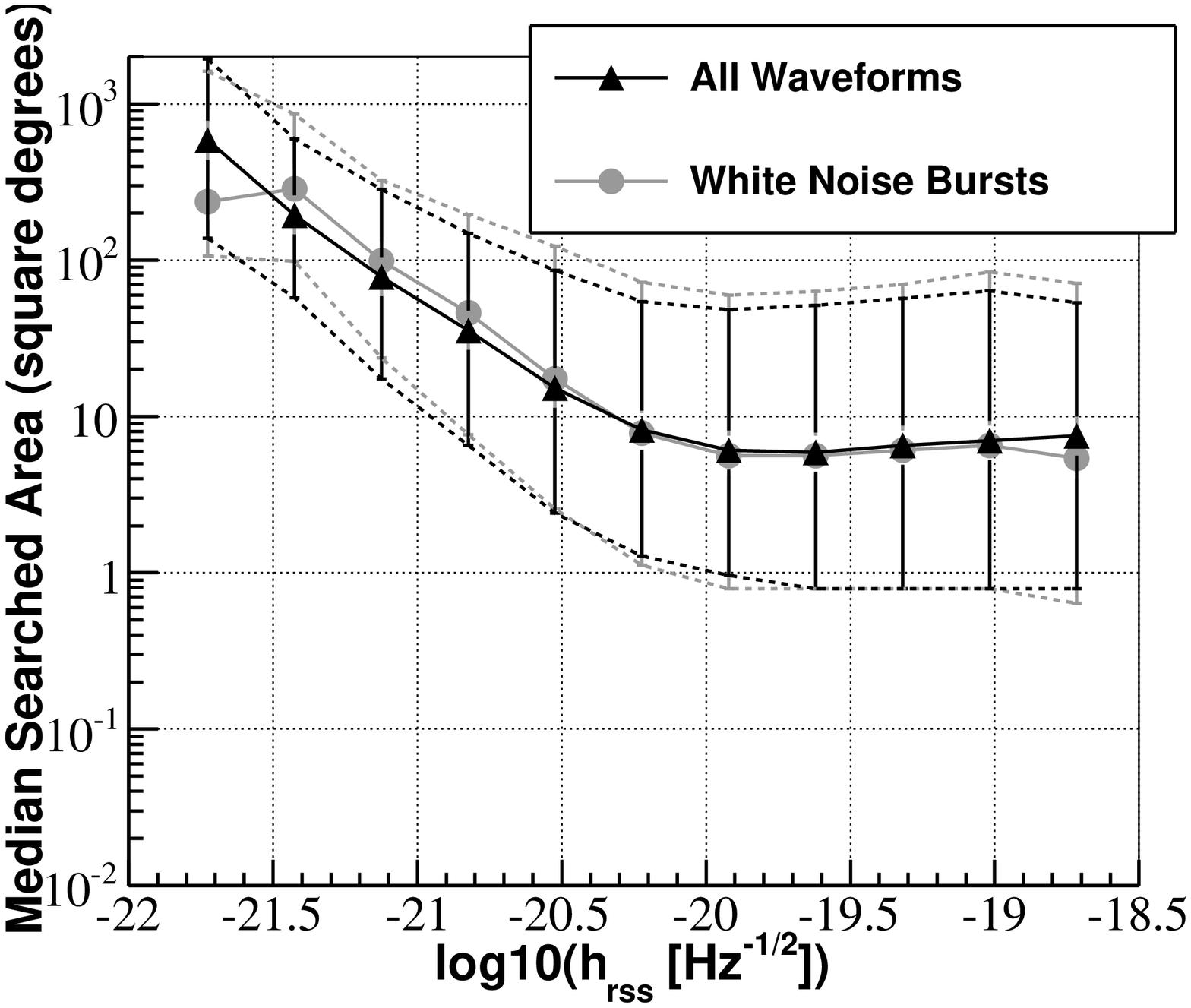} 
\hspace{0.5cm}
\includegraphics*[width=0.38\textwidth]{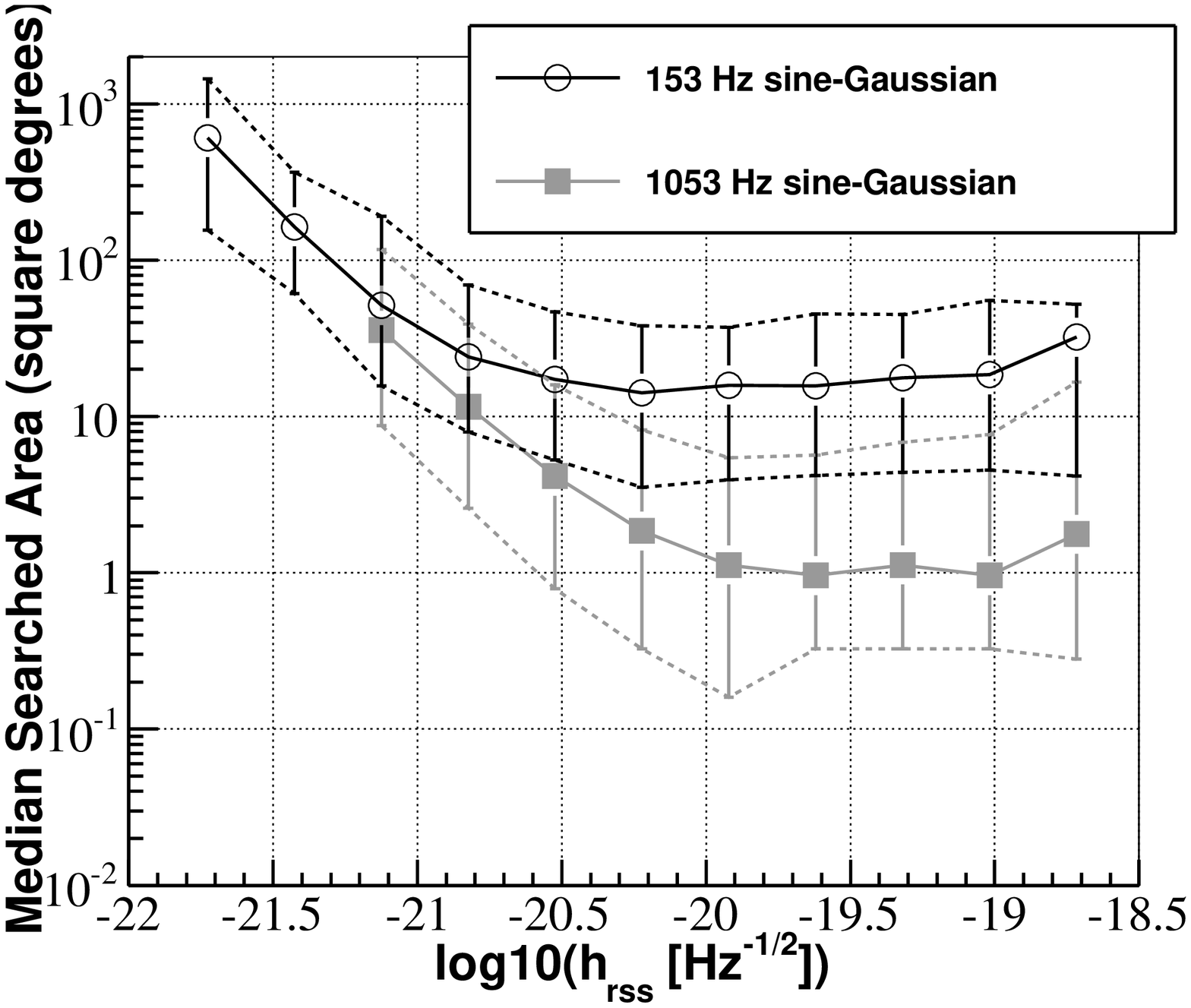}
}
\caption{Plots of typical uncertainty region sizes for the Omega (top)
and unconstrained cWB (bottom) pipelines, as a function of GW strain
amplitude at Earth, for various waveform types.
The ``searched area'' is the area of the skymap with a likelihood value
greater than the likelihood value at the true source location {\em before}
the galaxy catalog is used to further limit the search region.
The solid line with symbols represents
the median (50\%) performance, while the upper and lower dashed lines show
the 75\% and 25\% quartile values.
Near the detection threshold ($h_\text{rss} \sim 10^{-21} \,\text{Hz}^{-1/2}$) , uncertainty
regions are typically between 10 and 100 
square degrees.  The Omega pipeline performs poorly on white noise bursts but
exceptionally well on sine-Gaussians because it is designed to identify signals that are well-localized in frequency space.
}
\label{fig:SearchArea}
\end{center}
\ifthenelse{\equal{\targetjournal}{aa}}{\end{figure*}}{\end{figure*}}

An ensemble of simulated GW signals was generated to measure the
effectiveness of the reconstruction and follow-up procedures.  
For the Omega and cWB burst pipelines, these ``software injections'' were a mix of ad hoc sine-Gaussian, Gaussian, and white
noise burst waveforms similar in type and distribution to those used in
previous LIGO/Virgo all-sky analyses \citep{Abbott2009,Abadie2010}.  
While these waveforms are not based on specific astrophysical models, they do
a good job of characterizing detector response for signals in specific
frequency ranges (sine-Gaussians) and broadband signals (white-noise
bursts).  For MBTA (see Sect. \ref{trig_mbta}), injections were drawn from NS-NS and NS-BH inspiral waveforms
with a range of parameters.
To emulate
a realistic spatial distribution, each injection was calculated with
a source distance and direction inside
a randomly selected galaxy from the
GWGC and the simulated GW amplitudes were weighted to be inversely proportional to distance.  Only galaxies within 50 Mpc were included in the simulation, with weighting factors applied so that the probability of originating from
each galaxy was proportional to its blue light luminosity.
The simulation set and the analysis used the same catalog, so the results presented  
in Figures \ref{fig:cwb_astro} -- \ref{fig:cal} make the
assumption that the blue light luminosity distribution of galaxies
in the GWGC is a good tracer of GW sources in the local universe.
Signals
were superimposed on real LIGO-Virgo gravitational wave data taken
between August and December 2009.

While performance studies in this paper were done using software injections,
a relatively small number of tests in which a signal was physically put into the interferometer via actuators (``hardware injections'') were also performed, 
providing an additional cross-check.

\subsection{Testing Results}

Because the skymap likelihood regions are often irregularly shaped, the 
size of the uncertainty region is characterized by the ``searched area'', 
defined as the angular area of the skymap with likelihood greater than the likelihood at the 
true source location. The median searched area as a function of signal strength is plotted for both
cWB and Omega Pipeline in Fig. \ref{fig:SearchArea}.  Here, gravitational wave 
amplitudes are expressed in terms of their
root-sum-squared amplitude:

\begin{equation}
\label{eqn:hrss}
h_\text{rss} \equiv \sqrt{\int{\left(|h_+\left(t\right)|^2 + |h_{\times}\left(t\right)|^2\right)}\rm{dt}}
\end{equation}

\noindent where $h_+\left(t\right)$ and $h_{\times}\left(t\right)$ are the plus-
and cross-polarization strain functions of the wave.  Since $h$ is a 
dimensionless quantity,  $h_{\rm rss}$ is given in units of Hz$^{-1/2}$.
For this data, signals
near the detection threshold would have $h_\text{rss} \sim 10^{-21} \,\text{Hz}^{-1/2}$,
roughly corresponding to the cWB statistic $\eta \sim 5$ \citep{Abadie2010}.  These signals were
typically localized with median search areas of several tens of square degrees.  
The coherent position reconstruction algorithms are ``tuned'' to localize these
near-threshold signals as accurately as possible; as a result, some of the plots reveal 
a degradation in algorithm performance for very loud signals.
Median searched area is shown for MBTA in Fig. \ref{fig:MbtaArea}, as a function
of the combined SNR of the signal:

\begin{figure}
\begin{center}
\mbox{
\includegraphics*[width=0.8\columnwidth]{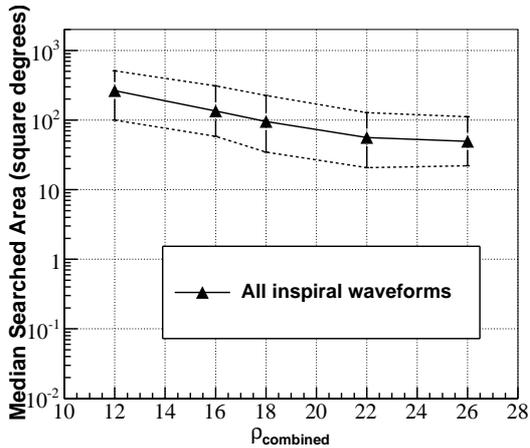}}
\caption{Plots of uncertainty region sizes for the MBTA pipeline as a
function of combined SNR ($\rho_{\rm{combined}}$).
The solid line with symbols represents
the median (50\%) performance, while the upper and lower dashed lines show
the 75\% and 25\% quartile values.  The expected detection threshold is around
$\rho_{\rm {combined}}\sim 12$.}
\label{fig:MbtaArea}
\end{center}
\end{figure}

\begin{equation}
\rho_{\mathrm {combined}} \equiv \sqrt{\rho_{\mathrm H1}^2+\rho_{\mathrm
L1}^2+\rho_{\mathrm V1}^2},
\end{equation}
where $\rho_{\mathrm H1}^2$, $\rho_{\mathrm L1}^2$, and $\rho_{\mathrm V1}^2$
are the single detector SNRs seen in the Hanford, Livingston and Virgo
instruments, respectively.

The simulated GW signals described above were also
used to test the tiling software in order to determine the success rate
for imaging the correct sky location with realistic detector noise,
reconstructed skymap shapes, and telescope FOVs.
The LUMIN software package was used to determine pointings for ground-based
telescopes and GEM was used for Swift. 

Some of the results of this simulation can be seen in Fig. \ref{fig:cwb_astro}.  
The results are plotted as a function of the 
ranking statistic used by each pipeline.  
On the y-axis, the
``Fraction of triggers imaged'' represents the fraction of triggers with the 
given detection statistic that have the correct image location included 
within the selected tiles.  Given a GW trigger, the success rate plotted in Fig. \ref{fig:cwb_astro}
estimates the odds of choosing
the right sky position.  In this figure, note that the ``whole skymap'' is 
limited to 160 square degrees, and so does not always include the true source 
location. The thresholds for initiating follow-ups varied with the 
condition of the interferometers, but was typically around 3.0 for
$\Omega$, $\eta = 3.5$ for cWB, and $\rho_{\rm{combined}} = 10$ for MBTA.
The complex behavior of the Omega efficiency curve is related to the
 use of a hybrid detection statistic which utilizes different methods 
depending on SNR range.  
Clearly, events with SNR near the threshold for triggering follow-ups, 
the most likely scenario for the first detections,
are the most difficult to localize.

\ifthenelse{\equal{\targetjournal}{aa}}{\begin{figure*}}{\begin{figure*}}
\begin{center}
\mbox{
\includegraphics[width=0.4\textwidth]{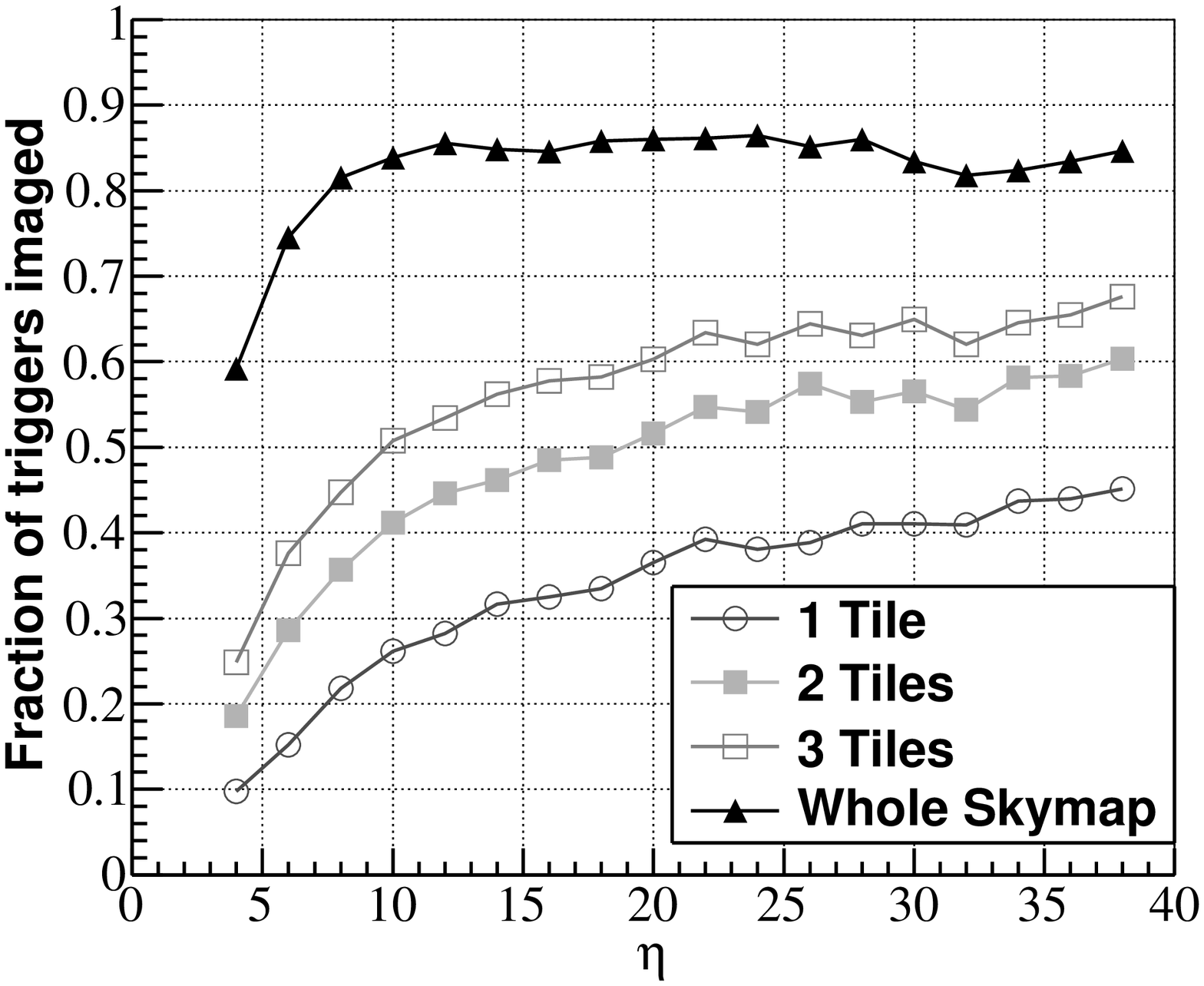} 
\includegraphics[width=0.4\textwidth]{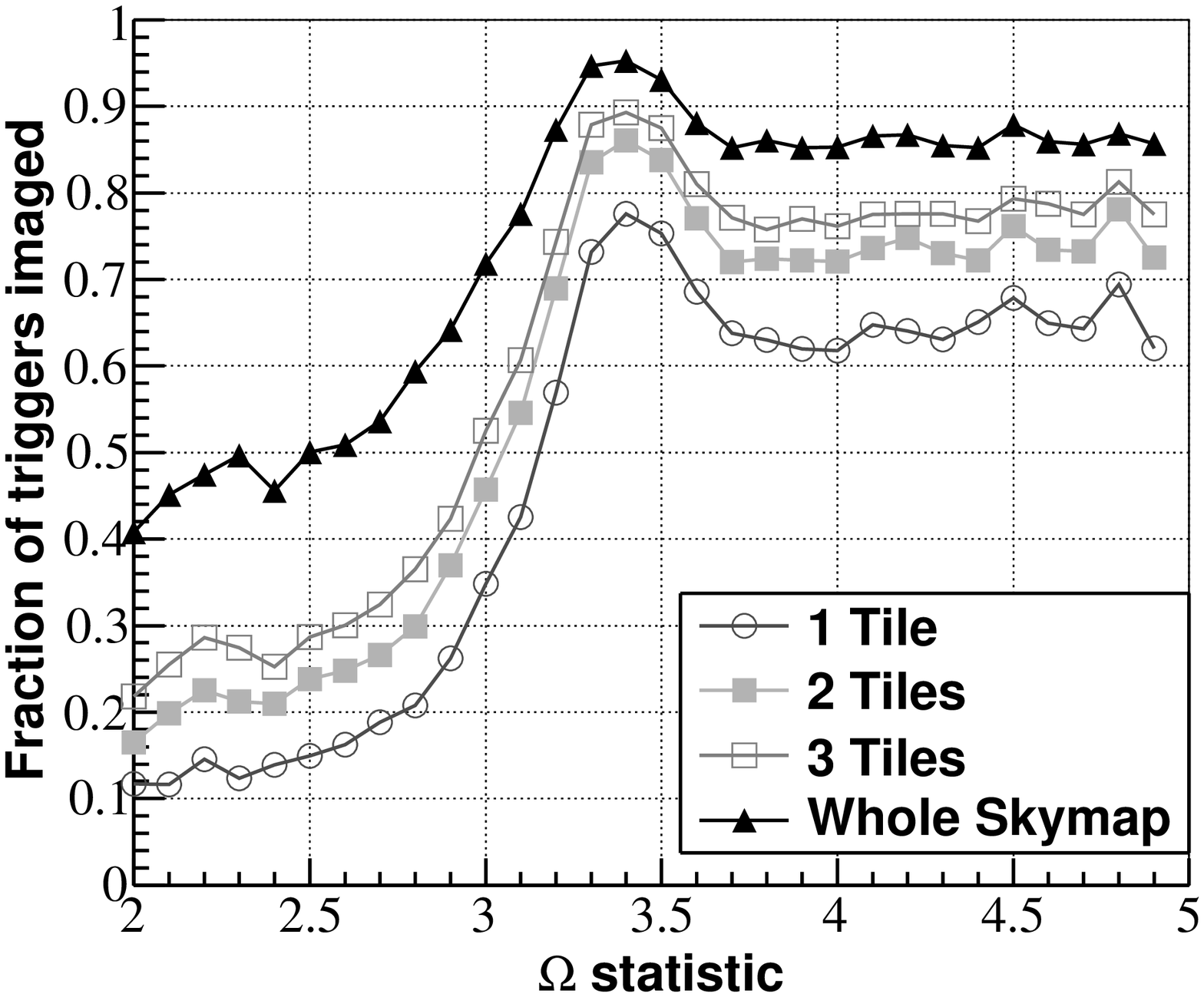}}
\mbox{
\includegraphics[width=0.4\textwidth]{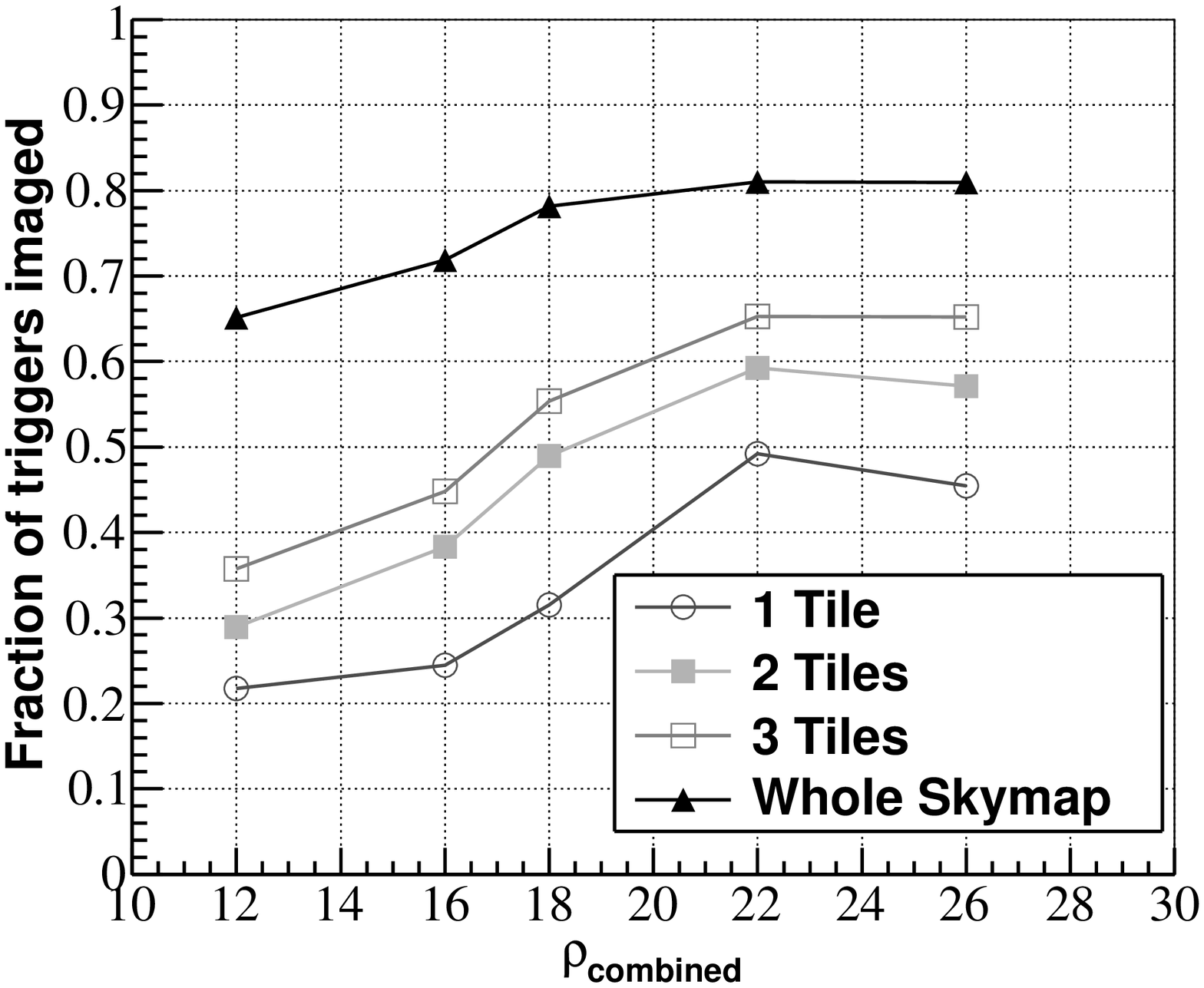}}
\caption{Success rates for the tile selection process based on unconstrained 
cWB (left), Omega (right), and MBTA (bottom) skymaps.  An injection recovered with the 
detection statistic 
shown on the horizontal axis is considered a success if the correct 
source location is included in one of the chosen tiles.  Typical thresholds
for follow-up are $\Omega$=3.0, $\eta$=3.5, and $\rho_{\rm{combined}}$=10.  
Each  tile is $1.85^{\circ} \times 1.85^{\circ}$, the FOV of both the ROTSE 
and TAROT telescopes. Statistical uncertainties are small with respect to the 
markers.}

\label{fig:cwb_astro}
\end{center}
\ifthenelse{\equal{\targetjournal}{aa}}{\end{figure*}}{\end{figure*}}

Example efficiency curves for the burst simulation are shown in 
Fig. \ref{fig:hrss}.  The efficiency for each marker on the plot is 
calculated as the fraction of signals for which the injected location was successfully imaged, 
for an
$h_{\rm{rss}}$ range centered on the marker. 
Specifically, we require that:

\begin{enumerate}

\item The trigger's ranking statistic is
higher than the threshold, which is chosen to enforce a FAR 
of about 1 GW trigger per day of livetime.

\item The true source location is included in one of the chosen
tiles.  Five tiles are allowed for Swift, three tiles for the 
QUEST camera, and one tile for all other telescopes.

\end{enumerate}

\noindent Note that efficiencies
in this figure do not
reach unity even for loud events primarily due to the difficulty of correctly
localizing GWs in some regions of the sky where the
antenna response of one or more interferometers is poor.

The efficiencies produced with these criteria are upper limits
on what would be detected in a real search. 
They assume
that the EM transient is very bright, and will always be detected if the 
correct sky location is imaged.  
The quoted efficiencies do not account for the fact that some 
chosen tiles will not be observed due to restrictions from weather, instrument 
availability, proximity to the Sun or Moon, or the application of a manual veto.
The exact behavior of the efficiency
curves will vary depending on the morphologies of the simulation waveforms
selected.  Finally, the chosen GW trigger FAR of one event per day presumes
the false alarm fraction from the EM transient classification pipeline will be low enough
to make a coincidence significant.   Many of these additional complications
and their associated impact on efficiency are described in
\citet{metzberg}. 

Nevertheless, these curves provide a measure of the potential for joint
EM/GW searches.  If the number of incidental EM transients in the observed
fields can be understood and controlled, then the addition of 
EM data can effectively increase the search sensitivity to very weak GW signals.  
For occasional strong GW signals, the plots suggest that only a few pointings of a 
telescope are enough to image the true location with better than 50\%
efficiency. 

\ifthenelse{\equal{\targetjournal}{aa}}{\begin{figure*}}{\begin{figure*}}
\begin{center}
\mbox{
\includegraphics[width=0.35\textwidth]{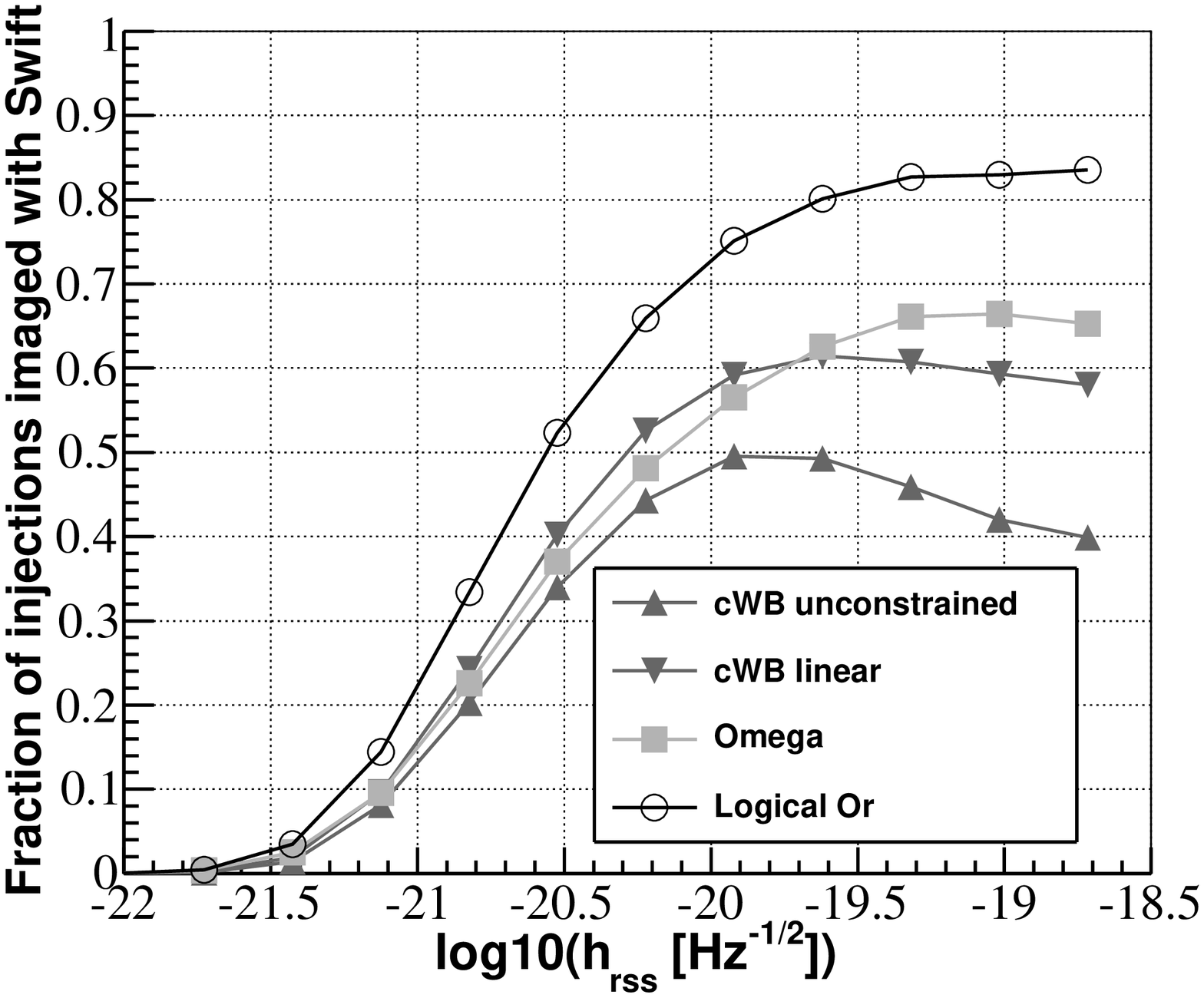}
\includegraphics[width=0.35\textwidth]{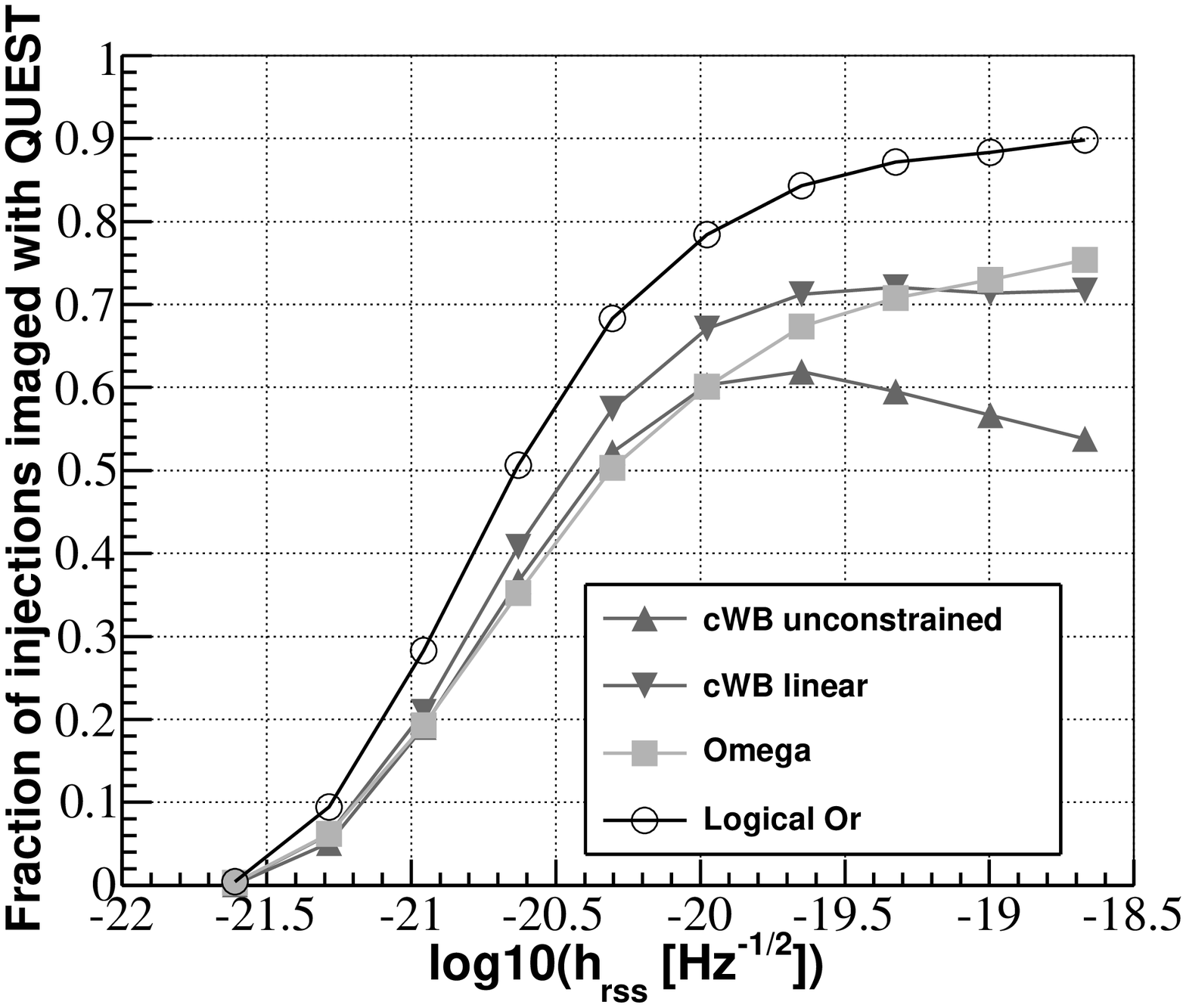}}
\caption{Fractional success as a function of strain at Earth for combinations
of the Omega and cWB burst pipelines.  Success rates assume
5 pointings for each event for Swift (left) and 3 for QUEST (right).
Fractional success
is end-to-end from triggering by pipeline to successful pointing, with a
threshold for follow-up approximating a FAR of one per day. The ``Logical Or''
curve counts a success if either linear cWB or Omega correctly localized the
event, effectively doubling the allowed number of tiles. 
Some curves show degraded performance for very loud signals because the 
algorithms are tuned to optimize performance for weaker events close to the detection
threshold.
Statistical uncertainties are small with respect to the markers.}

\label{fig:hrss}
\end{center}
\ifthenelse{\equal{\targetjournal}{aa}}{\end{figure*}}{\end{figure*}}

\subsection{Calibration Uncertainty}

\ifthenelse{\equal{\targetjournal}{aa}}{\begin{figure*}}{\begin{figure*}}
\begin{center}
\includegraphics[width=0.4\textwidth]{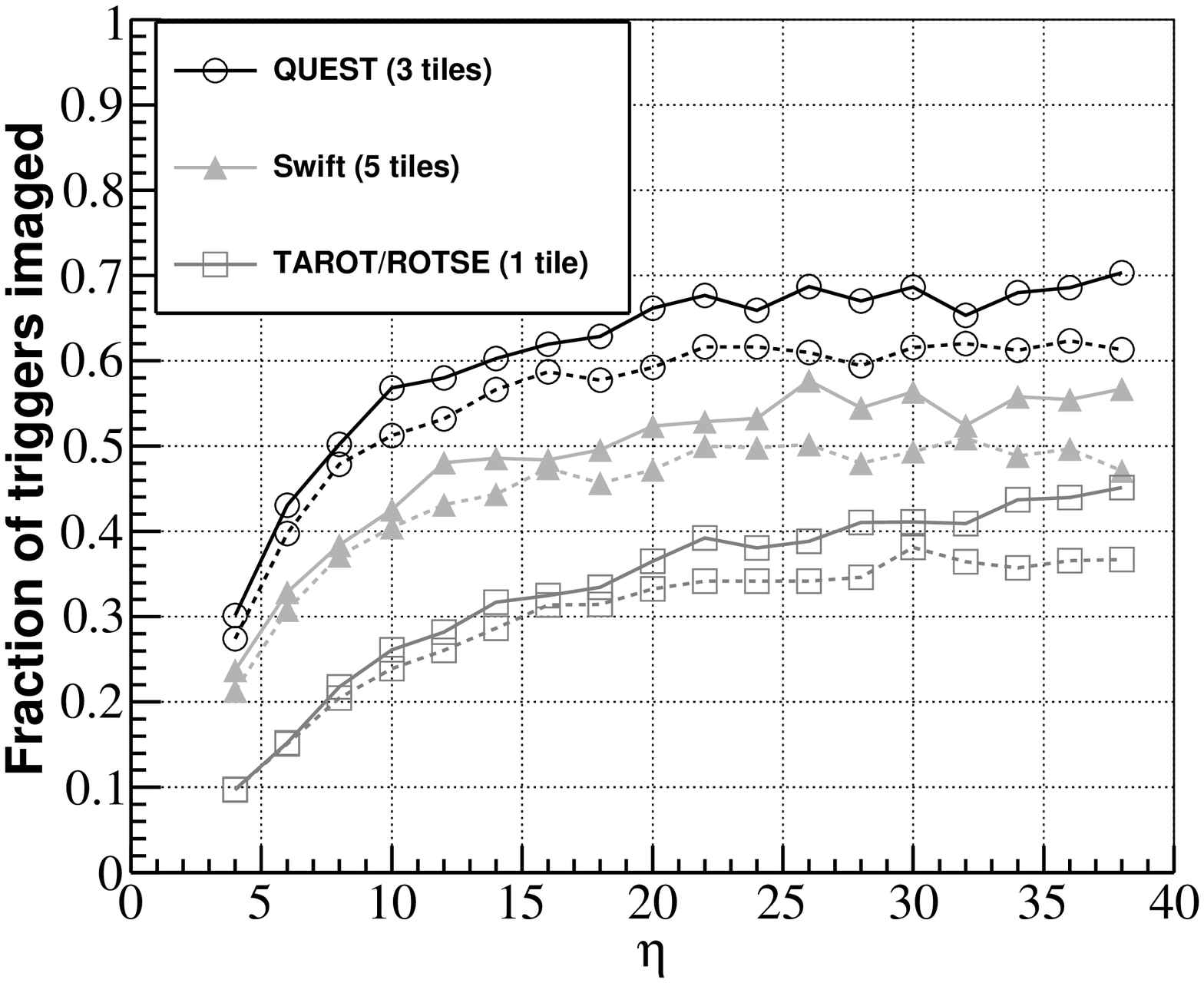}
\caption{Fractional success rate for simulated gravitational wave signals with
(dashed) and without (solid) including calibration uncertainties.  QUEST
 results assume 3 pointings, Swift results assume 5 pointings and TAROT/ROTSE results 
assume 1 pointing.}
\label{fig:cal}
\end{center}
\ifthenelse{\equal{\targetjournal}{aa}}{\end{figure*}}{\end{figure*}}

Uncertainty in the calibration of GW detectors  \citep{s5cal, virgocal}  may impact 
the ability to correctly choose the right fields to observe with EM
instruments.  To estimate the potential detriment to pointing, we generated 
a second set of simulated burst signals, with each signal including some level
of miscalibration corresponding to realistic calibration errors.  Before being 
added to detector noise, each astrophysical signal 
was scaled in amplitude by a factor between 
$0.85$ and $1.15$, and shifted in time by between $-150$ and $150$ $\mu$s.  
The exact amplitude and time ``jitter'' were randomly selected from flat 
distributions for each signal entering each detector.  The bounds of the 
distribution of values for the timing and amplitude jitter were chosen to match preliminary
estimates for the LIGO and Virgo calibration error budgets 
around 150 Hz for the 2009--2010 run.  
Well above this frequency, the actual timing errors are likely
less than this model; the simulation is conservative in 
this sense.

Some of the results of this simulation, with the cWB algorithm, 
may be seen in Fig. \ref{fig:cal}.
The success rate is shown for the entire pipeline,
assuming one pointing of a $1.85^\circ \times 1.85^\circ$ FOV, three pointings of the QUEST
FOV, and five pointings of a Swift FOV.  The curves are shown both with and
without the effects of calibration uncertainty.  For the low SNR signals that
are the most likely for first detections, $\eta \lesssim 10$, the efficiency
is within a few percent  with and without the calibration uncertainty.  This 
is expected, since at low signal to noise ratio, timing uncertainty 
from detector noise is larger than timing uncertainty due to 
calibration \citep{fairhurst}.  However, for louder signals, the ability 
to correctly choose the right sky location is seen to be modestly impacted 
by the accuracy of the calibration.

\section{Summary}\label{conclusions}

Mergers of compact binary systems containing neutron stars, as well as 
some other energetic astrophysical events, are expected
to emit observable transients in both the gravitational wave and electromagnetic
channels.  Observing populations of joint signals 
would likely reveal many details of the GW sources,
and could even constrain cosmological models.

During 2009 and 2010, the LIGO and Virgo
collaborations partnered with a large, heterogeneous  
group of EM observatories to jointly seek transient
signals.  X-ray, optical, and radio observatories collected
follow-up observations to GW triggers that were delivered 
with $\sim$30 minutes of latency.  Analysis
of the multi-instrument data set is currently in progress,
and the results of the search for jointly observed transients
will be published at a later date.

A Monte Carlo study of the GW data analysis algorithms used in
the low latency pipeline demonstrated
the ability of the LIGO/Virgo network to localize transient GW events 
on the sky.  Localization ability depends strongly on the SNR of the GW 
signal; lower SNR signals are more difficult to localize, but are
also the more likely scenario for the first detections.  Signals
with SNR near the detection threshold were localized with median
sky areas between 10 and 100 square degrees.  After limiting 
the search to known galaxies and Milky Way globular clusters
within the detection range of the GW
observatories, the simulation shows that the correct location of 
signals detected near threshold
can be imaged with 30-50\% success with three fields of size $1.85^\circ
\times 1.85^\circ$, for instance.
Moreover, the ability to image the source position is seen to
be only marginally impacted by realistic levels of calibration uncertainty.  

This search establishes a baseline for low-latency analysis with the next-generation GW detectors
Advanced LIGO and Advanced Virgo.
Installation of these second-generation
detectors is already in progress, with observations
expected to begin around 2015. 
Developing a low-latency response
to GW triggers represents the first steps
toward solving the many logistical and technical challenges that must be overcome
to collect prompt, multiwavelength, EM observations of GW source progenitors.
The integration of GW and EM observatories is likely to continue to develop
over the next few years as the scientific community prepares to utilize
the many opportunities promised by the impending global network of 
advanced GW detectors.

\ifthenelse{\equal{\targetjournal}{aa}}{
  \begin{acknowledgements}\label{sec:acknowledgements}

The authors gratefully acknowledge the support of the United States
National Science Foundation for the construction and operation of the
LIGO Laboratory, the Science and Technology Facilities Council of the
United Kingdom, the Max-Planck-Society, and the State of
Niedersachsen/Germany for support of the construction and operation of
the GEO600 detector, and the Italian Istituto Nazionale di Fisica
Nucleare and the French Centre National de la Recherche Scientifique
for the construction and operation of the Virgo detector. The authors
also gratefully acknowledge the support of gravitational wave research by these
agencies and by the Australian Research Council, 
the International Science Linkages program of the Commonwealth of Australia,
the Council of Scientific and Industrial Research of India, 
the Istituto Nazionale di Fisica Nucleare of Italy, 
the Spanish Ministerio de Educaci\'on y Ciencia, 
the Conselleria d'Economia Hisenda i Innovaci\'o of the
Govern de les Illes Balears, the Foundation for Fundamental Research
on Matter supported by the Netherlands Organisation for Scientific Research, 
the Polish Ministry of Science and Higher Education, the FOCUS
Programme of Foundation for Polish Science,
the Royal Society, the Scottish Funding Council, the
Scottish Universities Physics Alliance, The National Aeronautics and
Space Administration, the Carnegie Trust, the Leverhulme Trust, the
David and Lucile Packard Foundation, the Research Corporation, and
the Alfred P. Sloan Foundation.
The authors acknowledge support for TAROT from the French
Minist\`ere des Affaires \'Etrang\`eres and Minist\`ere de
l'Enseignement Sup\'erieur et de la Recherche.
The observations by ROTSE-III were supported by NASA grant 
NNX08AV63G and NSF grant PHY-0801007.
The work with Swift was partially supported through a NASA 
grant/cooperative agreement number NNX09AL61G to the Massachusetts 
Institute of Technology.
The contribution from the ``Pi of the Sky'' group was financed by the Polish 
Ministry of Science in 2008-2011 as a research project. 
We thank Joshua S. Bloom for useful discussions on the rates of PTF transients and their classification.
This document has been assigned LIGO Laboratory document number \ligodoc.
  \end{acknowledgements}
}{
  \begin{acknowledgments}\label{sec:acknowledgements}

The authors gratefully acknowledge the support of the United States
National Science Foundation for the construction and operation of the
LIGO Laboratory, the Science and Technology Facilities Council of the
United Kingdom, the Max-Planck-Society, and the State of
Niedersachsen/Germany for support of the construction and operation of
the GEO600 detector, and the Italian Istituto Nazionale di Fisica
Nucleare and the French Centre National de la Recherche Scientifique
for the construction and operation of the Virgo detector. The authors
also gratefully acknowledge the support of gravitational wave research by these
agencies and by the Australian Research Council, 
the International Science Linkages program of the Commonwealth of Australia,
the Council of Scientific and Industrial Research of India, 
the Istituto Nazionale di Fisica Nucleare of Italy, 
the Spanish Ministerio de Educaci\'on y Ciencia, 
the Conselleria d'Economia Hisenda i Innovaci\'o of the
Govern de les Illes Balears, the Foundation for Fundamental Research
on Matter supported by the Netherlands Organisation for Scientific Research, 
the Polish Ministry of Science and Higher Education, the FOCUS
Programme of Foundation for Polish Science,
the Royal Society, the Scottish Funding Council, the
Scottish Universities Physics Alliance, The National Aeronautics and
Space Administration, the Carnegie Trust, the Leverhulme Trust, the
David and Lucile Packard Foundation, the Research Corporation, and
the Alfred P. Sloan Foundation.
The authors acknowledge support for TAROT from the French
Minist\`ere des Affaires \'Etrang\`eres and Minist\`ere de
l'Enseignement Sup\'erieur et de la Recherche.
The observations by ROTSE-III were supported by NASA grant 
NNX08AV63G and NSF grant PHY-0801007.
The work with Swift was partially supported through a NASA 
grant/cooperative agreement number NNX09AL61G to the Massachusetts 
Institute of Technology.
The contribution from the ``Pi of the Sky'' group was financed by the Polish 
Ministry of Science in 2008-2011 as a research project. 
We thank Joshua S. Bloom for useful discussions on the rates of PTF transients and their classification.
This document has been assigned LIGO Laboratory document number \ligodoc.
  \end{acknowledgments}
}

\ifthenelse{\equal{\targetjournal}{aa}}{
 \bibliographystyle{aa}
}{
 \bibliographystyle{apj}
}

\end{twocolumn}

\clearpage
\onecolumn
\begin{table}
\caption{Partner Instrument Characteristic Properties}
\label{table:1}
\centering
\begin{supertabular}{c c c c c c}
\hline\hline
Name & Band & FOV (square degrees) & Aperture (m) & Exposure Time (s) & Limiting Magnitude \\
\hline
Palomar Transient Factory & Optical & 7.3 & 1.2  & 60 & 20.5\\
Pi of the Sky & Optical & 400 & 0.072 & 10 & 11.5\\
QUEST & Optical & 9.4 & 1 & 60 & 20\\
ROTSE III & Optical & 3.4 & 0.45 & 20 & 17.5\\
SkyMapper & Optical & 5.7 & 1.35 & 110 & 21\\
TAROT & Optical & 3.4 & 0.25 & 180 & 17.5\\
Zadko Telescope & Optical & 0.15 & 1 & 180 & 20\\
Liverpool Telescope & Optical & 0.0058  & 2 & 3600 & 21\\
LOFAR & Radio & $\sim$25 & N/A & 14400 & N/A\\
Swift & X-ray & 0.15 & N/A  & 200-5000 & N/A\\
Swift & UV, Optical & 0.078 & 0.3 & 200-5000 & 24 \\
\hline
\end{supertabular}
\tablefoot{Some characteristics of instruments involved in the search.  The shown limiting magnitudes are estimates, assuming favorable observing conditions.}
\end{table}

\begin{table}
\caption{Partner Instrument Follow-up Information}
\label{table:2}
\centering
\begin{supertabular}{c c c c c}
\hline\hline
Name & Run & Tiles per Trigger & Target Alerts Per Week & Triggers Imaged \\
\hline
Palomar Transient Factory & Autumn & 10 & 1/3 & 1\\
Pi of the Sky & Autumn & 1 & 1 & 1\\
QUEST & Both & 3 & 1  & 5\\
ROTSE III & Autumn & 1 & 1 & 5\\
SkyMapper & Autumn & $\sim$9 & 1 & 3\\
TAROT & Both & 1 & 1 & 3\\
Zadko Telescope & Autumn & 5 & 1 & 2\\
Liverpool Telescope & Autumn & 1 & 1 & 1\\
LOFAR & Autumn & 1 & 1 & 2\\
Swift & Both & 5 & 1/4 & 2\\
\hline
\end{supertabular}
\tablefoot{Follow-up information for instruments involved in the search.  Each instrument participated in either the autumn run, or both the winter and autumn runs.  The column marked ``Tiles per Trigger'' shows how many different field locations the instrument attempted to observe for each accepted GW event candidate.  The ``Target Alerts Per Week'' column shows that alerts were sent to PTF and Swift at a lower rate than the other observatories.  The final column shows the number of GW event candidates for which each instrument collected data.}
\end{table}

\begin{twocolumn}
\end{twocolumn}

\end{document}